\documentclass{ws-rv975x65}
\usepackage[square]{ws-rv-van}    
\usepackage{ws-index}             
\makeindex
\newcounter{bean}

\begin{document}

\chapter{Shell-correction and orbital-free density-functional methods for 
finite systems}
\label{yann_ofm}

\author[Constantine Yannouleas and Uzi Landman]
{Constantine Yannouleas\footnote{Constantine.Yannouleas@physics.gatech.edu} and 
Uzi Landman\footnote{Uzi.Landman@physics.gatech.edu}}

\address{School of Physics, Georgia Institute of Technology,
Atlanta, Georgia 30332-0430, USA} 

\begin{abstract}
Orbital-free (OF) methods promise significant speed-up of computations based on 
density functional theory (DFT). In this field, the development of accurate 
kinetic-energy density functionals remains an open question. In this chapter we 
review the shell-correction method (SCM, commonly known as Strutinsky's averaging 
method) applied originally in nuclear physics and its more recent formulation in the
context of DFT [Yannouleas and Landman, {\it Phys. Rev. B\/} {\bf 48}, 8376 
(1993)]. We demonstrate the DFT-SCM method through its earlier applications to 
condensed-matter finite systems, including metal clusters, fullerenes, and metal 
nanowires. The DFT-SCM incorporates quantum mechanical interference effects and thus
offers an improvement compared to the use of Thomas-Fermi-type kinetic energy 
density functionals in OF-DFT. 
\end{abstract}

\date{1 May 2013}

\body

\section{Introduction}

\subsection{Preamble}
\label{preamble}

Often theoretical methods (in particular computational techniques) are developed  
in response to emerging scientific challenges in specific fields. The development 
of the shell correction method (SCM) by Strutinsky in the late 1960's \cite{stru}
was motivated by the observation of large nonuniformities 
(oscillatory behavior) exhibited by a 
number of nuclear properties as a function of the nuclear size. These properties 
included: total nuclear masses, nuclear deformation energies, and large distortions 
and fission barriers. While it was understood already that the total energy of 
nuclei can be decomposed into an oscillatory part and one that shows a slow 
``smooth'' variation as a function of size, 
Strutinsky's seminal contribution was to calculate 
the two parts from different nuclear models: the former from the nuclear shell model
and the latter from the liquid drop model. In particular, the 
calculation of the oscillatory part was enabled by employing an averaging method
that smeared the single particle spectrum associated with a nuclear model potential.
It is recognized that the Strutinsky procedure provides a method which ``reproduces 
microscopic results in an optimal way using phenomenological models'' 
\cite{bookring}; in the Appendix we describe an adaptation of the Strutinsky 
phenomenological procedure to metal clusters; we term this procedure as the 
semi-empirical (SE)-SCM.

In this chapter, we focus on our development in the early 1990's of the microscopic 
density-functional-theory (DFT)-SCM approach \cite{yl1}, where we have shown that 
the total energy of a condensed-matter finite system can be identified with 
the Harris functional [see Eq.\ (\ref{enhar})], with the shell correction [Eq.\ 
(\ref{densh})] being expressed through both the kinetic energy, $T_{sh}$, of this
functional [Eq.\ (\ref{tsh})] and the kinetic energy of an extended-Thomas-Fermi 
(ETF) functional expanded to fourth-order density gradients [see $T_{ETF}$ in 
Eq.\ (\ref{t4th})]. It is important to note 
that in our procedure an optimized input density is used in the Harris functional. 
This optimization can be achieved through a variational procedure [using an 
orbital-free (OF) energy functional, e.g., the ETF functional with 4th-order 
gradients] with a parametrized trial density profile [see Eq.\ (\ref{rho})], or
through the use [see Eq.\ (\ref{rhoalt})] of the variational principle applied to 
an orbital-free energy functional. (For literature regarding orbital-free
kinetic-energy functionals, see, e.g., 
Refs. \cite{kaxi05,perr94,madd94,tete92,tric09,wang00}.)
A similar optimization of an OF/4th-order-ETF 
density has been shown to be consistent with the Strutinsky averaging approach 
\cite{brack01}. Such 4th-order optimization of the input density renders rather 
ambiguous any direct (term-by-term) comparison between the method proposed by us 
and subsequent treatments, which extend the DFT-SCM to include 
higher-order shell-correction terms without input-density optimization 
\cite{ullm01,delc04} (see also Ref.\ \cite{tric04}). Indeed, the input-density 
optimization (in particular with the use of 4th-order gradients) minimizes 
contributions from higher-order shell corrections.
    
In light of certain existing similarities between the physics of nuclei and 
clusters (despite the large disparity in spatial and energy scales and the 
different origins of inter-particle interactions in these systems), 
in particular the finding of electronic shell effects in clusters 
\cite{knig84,deheer93,brack93,home,bookekardt}, 
it was natural to use the jellium  model 
in the early applications of the DFT-SCM to clusters. However, as noted \cite{yl1} 
already early on, ``the very good agreement between our results and those 
obtained via Kohn-Sham self-consistent jellium calculations suggested that it 
would  be worthwhile to extend the application of our method to more general 
electronic structure calculations extending beyond the jellium model, where the 
trial density used for minimization of the ETF functional could be taken as a 
superposition of site densities, as in the Harris method.'' Additionally, 
generalization of the DFT-SCM method to calculations of extended (bulk and surface)
systems appeared rather natural. Indeed, recent promising applications of DFT-SCM 
in this spirit have appeared \cite{wang07,wang08}. In this case, 
the term ``shell correction effects'' is also maintained, although 
``quantum interference effects'' could be more appropriate for extended systems.

\subsection{Motivation for finite systems}
\label{motiv}
 
One of the principle themes in research on finite systems (e.g., nuclei, atomic
and molecular clusters, and nano-structured materials) is the search for 
size-evolutionary patterns (SEPs) of properties of such systems and 
elucidation of the physical principles underlying such patterns \cite{seps}.

Various physical and chemical properties of finite systems exhibit SEPs,
including: 
\begin{list}%
{\arabic{bean}.}{\usecounter{bean}
\setlength{\itemsep}{0.0cm} \setlength{\parsep}{0.0cm} }

\item
Structural characteristics pertaining to atomic arrangements and particle
morphologies and shapes;
\item
Excitation spectra involving bound-bound transitions, ionization potentials
(IPs), and electron affinities (EAs);
\item
Collective excitations (electronic and vibrational);
\item
Magnetic properties;
\item
Abundance spectra and stability patterns, and their relation to 
binding and cohesion energetics, and to the pathways and rates of
dissociation, fragmentation, and fission of charged clusters;
\item
Thermodynamic stability and phase changes;
\item
Chemical reactivity.
\end{list}

\begin{figure}[t]
\centering\includegraphics[width=8.cm]{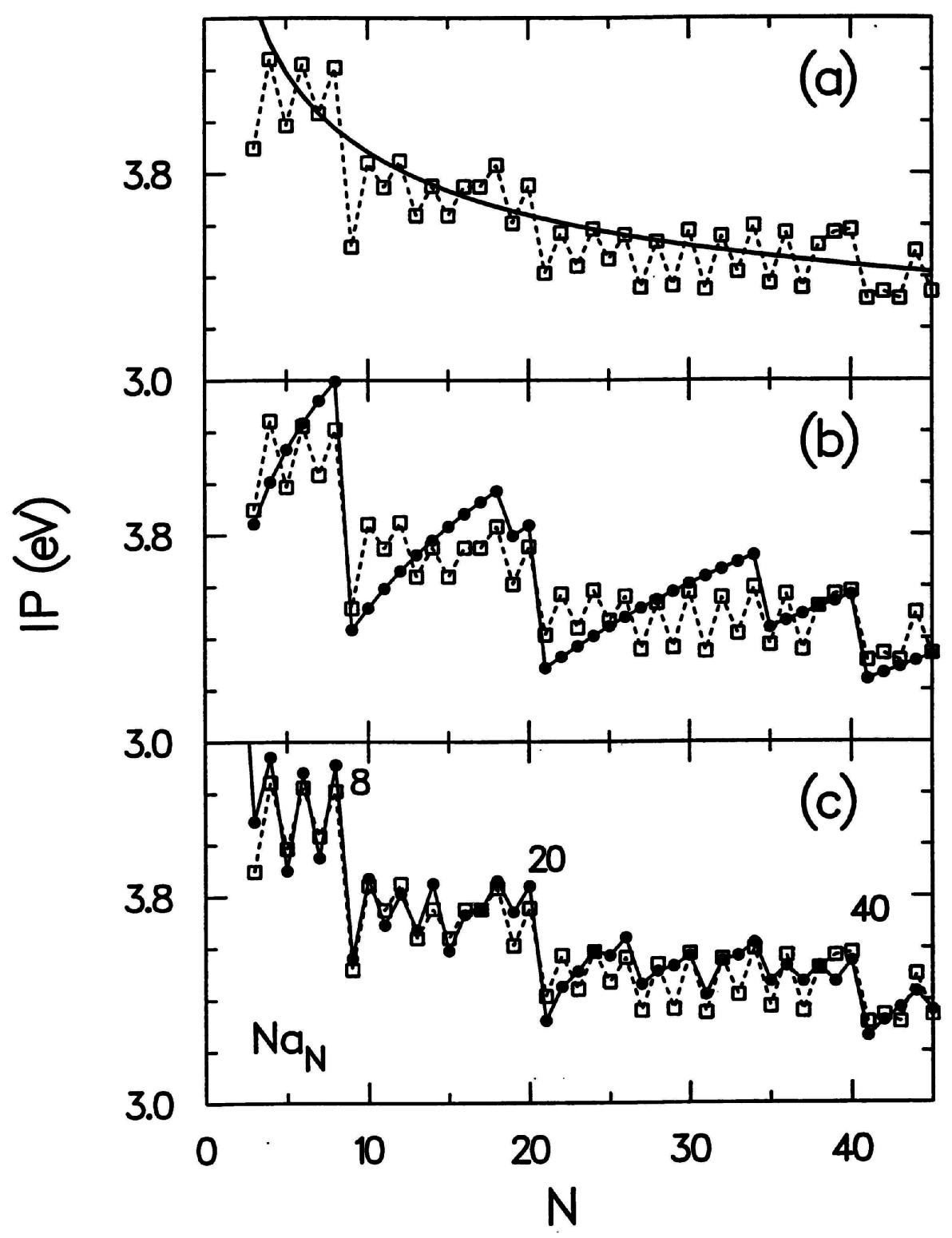}
\caption{Ionization potentials of Na$_N$ clusters. Open squares: Experimental 
measurements of Refs.\ \cite{deheer93,home}. 
The solid line at the top panel (a) represents 
the smooth contribution to the theoretical total IPs. The solid circles in the 
middle (b) and bottom (c) panels are the total SCM IPs. The shapes of sodium 
clusters have been assumed spherical in the middle panel, while triaxial 
deformations have been considered at the bottom one.
}
\label{fig1}
\end{figure}

The variations with size of certain properties of materials aggregates are commonly
found to scale on the average with the surface to volume ratio of the cluster,
i.e., $S/\Omega \sim R^{-1} \sim N^{1/3}$, where $S$, $\Omega$, $R$, and
$N$ are the surface area, volume, average radius, and number of particles,
respectively (the physical origins of such scaling
may vary for different properties).  In general, the behavior of SEPs in
finite systems in terms of such scaling is non-universal, in the sense that it
is non-monotonic exhibiting characteristic discontinuities. Nevertheless, in many 
occasions, it is convenient to analyze the energetics of finite systems in terms of
two contributions, namely, (i) a term which describes the energetics as a function
of the system size in an average sense (not including shell-closure
effects), referrred to usually as describing the ``smooth" part of the size 
dependence, and (ii) an electronic shell-correction term. The first term
is the one which is expected to vary smoothly and be expressible as an
expansion in $S/\Omega$, while the second one contains the characteristic
oscillatory patterns as the size of the finite system is varied. 
Such a strategy has been introduced \cite{stru} and often used in studies of nuclei
\cite{bookring,pres}, and has been adopted recently for investigations
of metal clusters \cite{yl1,yl2,yl3,yl4,yl5,yl6,suga,bulg,frau,reim,%
yann97.3,yann97,yann99,yann00,yann01,yann01.2,yann02}, fullerenes \cite{yl7}, 
and metal nanowires \cite{yann97.2,yann98,staff97}. As a motivating example, 
we show in \fref{fig1} the SEP of the IPs of Na$_N$ clusters, 
which illustrates odd-even oscillations in the observed spectrum, a smooth
description of the pattern [\fref{fig1}(a)], and two levels of shell-corrected
descriptions --- one assuming spherical symmetry [\fref{fig1}(b)], and the other 
allowing for triaxial shape deformations [\fref{fig1}(c)]. The progressive
improvement of the level of agreement between the experimental and theoretical
patterns is evident.

\subsection{Plan of the chapter}

The chapter is organized as follows:

In \sref{scm}, the general methodology of shell-correction methods is 
reviewed, and the microscopic DFT-SCM is introduced and presented in detail in
\sref{dftscm}.

Applications of the DFT-SCM to condensed-matter finite systems are presented
in \sref{appl} for three different characteristic nanosystems, namely, metal 
clusters (\sref{metclu}), charged fullerenes (\sref{full}), and metallic 
nanowires (\sref{wire}).

In the Appendix, we describe the semiempirical SCM for clusters,
which is closer to the spirit of Strutinsky's original phenomenological
approach for nuclei. There we also briefly present applications of the SE-SCM
to triaxial shape deformations and fission of metal clusters.

A summary is given in \sref{summ}.

\section{Methodology and derivation of microscopic DFT-SCM}
\label{scm}

\subsection{Historical review of SCM}

It has long been recognized in nuclear physics that the dependence 
of ground-state properties of nuclei on the number of particles 
can be viewed as the sum of two contributions: the first contribution varies 
smoothly with the particle number (number of protons $N_p$ and neutrons $N_n$)
and is referred to as the {\it smooth} part; the second contribution gives
a superimposed structure on the smooth curve and exhibits 
an oscillatory behavior, with extrema at the nuclear magic numbers
\cite{bm,pres}.

Nuclear masses have provided a prototype for this behavior \cite{bm}.
Indeed, the main contributions to the experimental nuclear binding energies 
are smooth functions of
the number of protons and neutrons, and are described by the 
semi-empirical mass formula \cite{weiz,beth}.
The presence of these smooth terms led to the introduction of 
the liquid-drop model (LDM), according to which the nucleus is viewed
as a drop of a nonviscous fluid whose total energy is specified by 
volume, surface, and curvature contributions \cite{bm,pres,mysw}.

The deviations of the binding energies from the smooth variation 
implied by the LDM have been shown \cite{mysw,stru} to arise 
from the shell structure associated with the bunching of the discrete 
single-particle spectra of the nucleons, and are commonly referred to as 
the shell correction.
Substantial progress in our understanding of the stability of strongly
deformed open-shell nuclei and of the dynamics of nuclear fission
was achieved when Strutinsky proposed \cite{stru} a physically motivated
efficient way of calculating the shell corrections. The method consists of
averaging [see the Appendix, \eref{espoc} and \eref{struav}] the single-particle
spectra of phenomenological deformed potentials and of
subtracting the ensuing average from the total sum of single-particle 
energies.

While certain analogies, portrayed in experimental data, 
between properties of nuclei and elemental clusters have been recognized, the 
nuclear-physics approach of separating the various quantities as a function 
of size into a smooth part and a shell correction part has 
only partially been explored in the case of metal clusters.
In particular, several investigations \cite{snso,brac2,serr,memb}
had used the ETF method in conjunction with the jellium approximation
to determine the average (smooth, in the sense defined above) behavior of metal
clusters, but had not pursued a method for calculating the shell corrections. 

In the absence of a method for appropriately calculating shell-corrections 
for metal clusters in the context of
the semiclassical ETF method, it had been presumed that the ETF method was most 
useful for larger clusters, since the shell effects diminish with 
increasing size. Indeed, several studies had been carried out with this method
addressing the asymptotic behavior of ground-state properties towards the
behavior of a jellium sphere of infinite size \cite{enge,seid}.

It has been observed \cite{brac2,yann,yann1,yann2,yann3}, however, that the 
single-particle potentials resulting from the semiclassical method are very 
close, even for small cluster sizes, to those obtained via self-consistent 
solution of the local density functional approximation (LDA) using
the Kohn-Sham (KS) equations \cite{ks}.
These semiclassical potentials were used
extensively to describe the optical (linear) response
of spherical metal clusters,
for small \cite{yann,yann1,yann2}, as well as larger sizes \cite{yann3}
(for an experimental review on optical properties, cf.\ Refs.\
\cite{dehe2,kres}). 
The results of this approach are consistent with time-dependent local density 
functional approximation (TDLDA) calculations which use the KS 
solutions \cite{ekar1,beck2}. 

It is natural to explore the use of these semiclassical potentials,
in the spirit of Strutinsky's approach, for evaluation of shell 
corrections in metal clusters of arbitrary size. Below we describe a
microscopic derivation of an SCM in conjunction with the density
functional theory \cite{yl1,yl2,yl3}, 
and its applications in investigations of the properties
of metal clusters and fullerenes. Particularly interesting
and promising is the manner by which the shell corrections are introduced by us 
at the microscopic level through the kinetic energy term \cite{yl1,yl2,yl3}, 
instead of the traditional semiempirical Strutinsky averaging procedure of the 
single-particle spectrum \cite{stru,reim}. In particular, our approach leads
to an energy functional that corrects many shortcomings of the orbital-free DFT,
and one that is competitive in numerical accuracy and largely advantageous in 
computational speed compared to the KS method.

\subsection{DFT-SCM}
\label{dftscm}

Underlying the development of the shell-correction method is the idea of
approximating the total energy $E_\mathrm{total} (N)$ of a finite interacting
fermion system as
\begin{equation}
E_\mathrm{total} (N) =  \widetilde{E} (N) + \Delta E_{{sh}} (N),
\label{enshsm}
\end{equation}
where $\widetilde{E}$ is the part that varies smoothly as a function of system
size, and $\Delta E_{{sh}}$ is an oscillatory term. Various
implementations of such a separation consist of different choices and methods
for evaluating the two terms in Eq.\ (\ref{enshsm}). Before discussing
such methods, we outline a microscopic derivation of Eq.\ (\ref{enshsm}).

Motivated by the behavior of the empirical nuclear binding
energies, Strutinsky conjectured that the self-consistent Hartree-Fock density 
$\rho_{{HF}}$ can be decomposed into a smooth density $\widetilde{\rho}$
and a fluctuating contribution $\delta \rho$, namely 
$\rho_{{HF}}=\widetilde{\rho}+\delta \rho$.
Then, he proceeded to
show that, to second-order in $\delta \rho$, the Hartree-Fock energy is 
equal to the result that the same Hartree-Fock expression yields
when $\rho_{{HF}}$ is replaced by the smooth density $\widetilde{\rho}$
and the Hartree-Fock single-particle energies $\varepsilon_i^{{HF}}$
are replaced by the single-particle energies corresponding to the smooth 
potential constructed with the smooth density $\widetilde{\rho}$. Namely, he 
showed that
\begin{equation}
E_{{HF}} = E_{{Str}} + O(\delta \rho^2),
\label{estr}
\end{equation}
where the Hartree-Fock electronic energy is given by the expression 
\begin{equation}
E_{{HF}} = \sum_{i=1}^\mathrm{occ} \varepsilon_i^{{HF}}
-\frac{1}{2} \int d\/{\bf r} d\/{\bf r^\prime} 
{\cal V} ({\bf r}-{\bf r^\prime})
[\rho_{{HF}}({\bf r},{\bf r})  
\rho_{{HF}}({\bf r^\prime},{\bf r^\prime}) -
\rho_{{HF}}({\bf r},{\bf r^\prime})^2 ],
\label{enhf}
\end{equation}
with $\varepsilon_i^{{HF}}$ being the eigenvalues obtained through a
self-consistent solution of the HF equation,
\begin{equation}
\left( -\frac{\hbar^2}{2m} \nabla^2 + U_{{HF}} \right)
\phi_i = \varepsilon_i^{{HF}} \phi_i,
\label{sphf}
\end{equation}
where 
\begin{equation}
U_{{HF}} ({\bf r}) \phi_i({\bf r})=
\int \! d\/ {\bf r^\prime} {\cal V} ({\bf r} - {\bf r^\prime})
[\rho_{{HF}} ({\bf r^\prime} , {\bf r^\prime}) \phi_i({\bf r}) -
\rho_{{HF}} ({\bf r^\prime} , {\bf r})\phi_i({\bf r^\prime})].
\label{pothf}
\end{equation}

The Strutinsky approximate energy is written as follows,
\begin{equation}
E_{{Str}} = \sum_{i=1}^\mathrm{occ} \widetilde{\varepsilon}_i
-\frac{1}{2} \int d\/{\bf r} d\/{\bf r^\prime} 
{\cal V} ({\bf r}-{\bf r^\prime})
[\widetilde{\rho} ({\bf r},{\bf r})  
\widetilde{\rho} ({\bf r^\prime},{\bf r^\prime}) -
\widetilde{\rho} ({\bf r},{\bf r^\prime})^2 ],
\label{enstr}
\end{equation}
where the index $i$ in Eq.\ (\ref{enhf}) and Eq.\ (\ref{enstr}) runs 
only over the occupied states (spin degeneracy is naturally implied).
The single-particle energies $\widetilde{\varepsilon}_i$ correspond to a
smooth potential $\widetilde{U}$. Namely, they are eigenvalues
of a Schr\"{o}dinger equation,
\begin{equation}
\left( -\frac{\hbar^2}{2m} \nabla^2 + \widetilde{U} \right)
\varphi_i = \widetilde{\varepsilon}_i \varphi_i,
\label{sphfsm}
\end{equation}
where the smooth potential $\widetilde{U}$ depends on the smooth
density ${\widetilde{\rho}}$, i.e.,
\begin{equation}
\widetilde{U}({\bf r}) \varphi_i({\bf r})=
\int \! d\/ {\bf r^\prime} {\cal V} ({\bf r} - {\bf r^\prime})
[\widetilde{\rho}({\bf r^\prime} , {\bf r^\prime}) \varphi_i({\bf r}) -
\widetilde{\rho}({\bf r^\prime} , {\bf r})\varphi_i({\bf r^\prime})],
\label{pothfsm}
\end{equation}
and ${\cal V}$ is the nuclear two-body interaction potential.

It should be noted that while Eq.\ (\ref{enstr})$-$ Eq.\ (\ref{pothfsm}) look 
formally
similar to the Hartree-Fock equations (\ref{enhf}-\ref{pothf}), their content
is different. Specifically, while in the HF equations, the density 
$\rho_{HF}$ is self-consistent with the wavefunction solutions of Eq.\ 
(\ref{sphf}), the density $\widetilde{\rho}$ in Eq.\ (\ref{enstr})$-$
Eq.\ (\ref{pothfsm}) is not self-consistent with the wavefunction
solutions of the corresponding single-particle Eq.\ (\ref{sphfsm}), i.e.,
$\widetilde{\rho} \neq \sum_{i=1}^\mathrm{occ} | \varphi_i |^2$. 
We return to this issue below.

Since the second term in Eq.\ (\ref{enstr}) is a smooth quantity,
Eq.\ (\ref{estr}) states that all shell corrections are, to first order in 
$\delta \rho$, contained in the sum of the single-particle energies 
$\sum_{i=1}^\mathrm{occ} \widetilde{\varepsilon}_i$. Consequently,
Eq.\ (\ref{enstr}) can be used as a basis for a separation of the total
energy into smooth and shell-correction terms as in Eq.\ (\ref{enshsm}).
Indeed Strutinsky suggested a semi-empirical 
method of such separation through an averaging
procedure of the single-particle energies $\widetilde{\varepsilon}_i$ in
conjunction with a phenomenological (or semi-empirical) model 
[the liquid drop model (LDM)] for the smooth part (see the appendix).

Motivated by the above considerations, 
we have extended them \cite{yl1,yl2,yl3} in the context of
density functional theory for electronic structure calculations. First we
review pertinent aspects of the DFT theory. In DFT, the total energy is given by 
\begin{equation}
E[\rho]=T[\rho]  + 
\int \left\{ \left[ \frac{1}{2} V_H [\rho({\bf r})]
+ V_I({\bf r}) \right] \rho({\bf r}) \right\}  d\/ {\bf r} + 
\int {\cal E}_{{xc}} [\rho({\bf r})]d\/ {\bf r} + E_I,
\label{enlda}
\end{equation}
where $V_H$ is the Hartree repulsive potential among the electrons, $V_I$ is
the interaction potential between the electrons and ions, 
${\cal E}_{{xc}}$ is the exchange-correlation functional [the
corresponding xc potential is given as $V_{{xc}}({\bf r}) \equiv
\delta {\cal E}_{{xc}} \rho({\bf r}) / \delta\rho({\bf r})$] and 
$T[\rho]$ is given in terms of a yet unknown functional 
$t[\rho({\bf r})]$ as $T[\rho] = \int t[\rho( {\bf r})] d{\bf r}$.
$E_I$ is the interaction energy of the ions.

In the Kohn-Sham (KS)-DFT theory, the electron density is evaluated from the
single-particle wave functions $\phi_{{KS,i}}({\bf r})$ as
\begin{equation}
\rho_{{KS}}({\bf r})= \sum_{i=1}^{\mathrm{occ}} 
\left| \phi_{{KS,i}}({\bf r}) \right|^2,
\label{rhoks}
\end{equation}
where $\phi_{{KS,i}}({\bf r})$ are obtained from a self-consistent solution
of the KS equations,
\begin{equation}
\left[ -\frac{\hbar^2}{2m} \nabla^2 + V_{{KS}} \right] 
\phi_{{KS,i}}({\bf r}) = \varepsilon_{{KS,i}} \phi_{{KS,i}}({\bf r})
\label{kseqs}
\end{equation}
where 
\begin{equation}
V_{{KS}} [\rho_{KS} ({\bf r})] = V_H [\rho_{KS} ({\bf r})]+
V_{{xc}}[\rho_{KS} ({\bf r})] + V_I({\bf r}). 
\label{VKS}
\end{equation}

The kinetic energy term in Eq.\ (\ref{enlda}) is given by 
\begin{equation}
T[\rho_{KS}] = \sum_{i=1}^\mathrm{occ} < \phi_{{KS,i}} | 
-\frac{\hbar^2}{2m} \nabla^2 |\phi_{{KS,i}}>,
\label{TKS}
\end{equation}
which can also be written as
\begin{equation}
T[\rho_{KS}] = \sum_{i=1}^\mathrm{occ} \varepsilon_{KS,i} -
\int \! \rho_{KS} ({\bf r}) V_{KS} [\rho_{KS} ({\bf r})] d{\bf r}.
\label{TKS2}
\end{equation}

According to the Hohenberg-Kohn theorem, the energy functional (\ref{enlda})
is a minimum at the true ground density $\rho_{gs}$, which in the context of
the KS-DFT theory corresponds the the density, $\rho_{KS}$, obtained from an
iterative self-consistent solution of Eq.\ (\ref{kseqs}). In other words,
combining Eq.\ (\ref{enlda}) and Eq.\ (\ref{TKS2}), and denoting by ``in" and 
``out" the trial and output densities of an iteration cycle in the solution
of the KS equation [Eq.\ (\ref{kseqs})], one obtains,
\begin{eqnarray}
E_{KS}[\rho_{KS}^\mathrm{out}] & = & E_I + \sum_{i=1}^\mathrm{occ} 
\varepsilon_{KS,i}^\mathrm{out} + \nonumber \\ 
~ & ~ & \int \! \left\{ \frac{1}{2} V_H [\rho_{KS}^\mathrm{out} ({\bf r})] +
{\cal E}_{xc} [\rho_{KS}^\mathrm{out} ({\bf r})] +V_I ({\bf r}) \right\} 
\rho_{KS}^\mathrm{out} ({\bf r}) d{\bf r} -  \nonumber \\
~& ~ & \int \! \rho_{KS}^\mathrm{out} ({\bf r}) 
V_{KS} [\rho_{KS}^\mathrm{in} ({\bf r})] d{\bf r}.
\label{eniter}
\end{eqnarray}
Note that the expression on the right involves both $\rho_{KS}^\mathrm{out}$ and
$\rho_{KS}^\mathrm{in}$.
Self-consistency is achieved when $\delta \rho_{KS}^\mathrm{out,in} ({\bf r}) =
\rho_{KS}^\mathrm{out} ({\bf r}) - \rho_{KS}^\mathrm{in} ({\bf r})$ becomes 
arbitrarily small (i.e., when $\rho_{KS}^\mathrm{out}$ converges to $\rho_{KS}$).

On the other hand, it is desirable to introduce approximate energy
functionals for the calculations of ground-state electronic properties,
providing simplified, yet accurate, computational schemes. It is indeed
possible to construct such functionals \cite{harr,finn,pome,foul,zare}, an example 
of which was introduced by J. Harris \cite{harr}, 
where self-consistency is circumvented and the result is accurate 
to second order in the difference between the trial and the self-consistent 
KS density {\bf (}see in particular Eq.\ (24a) of Ref.\ \cite{zare}; 
the same also holds true for the difference between the trial and the output
densities of the Harris functional{\bf )}.

The expression of the Harris functional is obtained from Eq.\ (\ref{eniter})
by dropping the label $KS$ and 
by replacing everywhere $\rho^\mathrm{out}$ by $\rho^\mathrm{in}$, yielding 
[note cancellations between the third and fourth terms on the right-hand-side
of Eq.\ (\ref{eniter})]. 
\begin{eqnarray}
E_\mathrm{Harris}[\rho^\mathrm{in}] & = & E_I + \sum_{i=1}^\mathrm{occ} 
\varepsilon_i^\mathrm{out} - 
 \int \! \left\{ \frac{1}{2} V_H [ \rho^\mathrm{in} ({\bf r})] +
V_{xc} [ \rho^\mathrm{in} ({\bf r})] \right\} \rho^\mathrm{in} ({\bf r}) d{\bf r} +
\nonumber \\
~& ~ & \int \! {\cal E}_{xc} [ \rho^\mathrm{in} ({\bf r})] d{\bf r}.
\label{enhar}
\end{eqnarray}
$\varepsilon_i^\mathrm{out}$ are the single-particle solutions (non-self-consistent)
of Eq.\ (\ref{kseqs}), with $V_{KS}[\rho^\mathrm{in} ({\bf r})$ [see Eq.\
(\ref{VKS})].

As stated above this result is accurate to second order in $\rho^\mathrm{in}
-\rho_{KS}$ (alternatively in $\rho^\mathrm{in}-\rho^\mathrm{out}$), thus
approximating the self-consistent total energy $E_{KS}[\rho_{KS}]$.

Obviously the accuracy of the results obtained via Eq.\ (\ref{enhar}) depend
on the choice of the input density $\rho^\mathrm{in}$. In electronic structure
calculations where the corpuscular nature of the ions is included 
(i.e., all-electron or pseudo-potential calculations), a natural choice for
$\rho^\mathrm{in}$ consists of a superposition of atomic site densities, 
as suggested originally by Harris. In the case of jellium calculations, we have 
shown \cite{yl1} that an accurate approximation to the KS-DFT total energy is
obtained by using the Harris functional with the input density, $\rho^\mathrm{in}$,
in Eq.\ (\ref{enhar}) evaluated from an Extended-Thomas-Fermi (ETF)-DFT 
calculation.

The ETF-DFT energy functional, $E_{ETF} [\rho]$, is obtained by replacing the
kinetic energy term in Eq.\ (\ref{enlda}) by a kinetic energy 
density-functional in the spirit of the Thomas-Fermi approach \cite{tf}, 
but comprising terms up to fourth-order in the density gradients 
\cite{brac2,hodg}. The optimal
ETF-DFT total energy is then obtained by minimization of $E_{ETF} [\rho]$
with respect to the density. In our calculations, we use for the trial
densities parametrized profiles $\rho ({\bf r};\; \{\gamma_i\})$ with
$\{ \gamma_i \}$ as variational parameters (the ETF-DFT optimal density is
denoted as $\rho_{ETF}$). The single-particle eigenvalues,
$\{\varepsilon_i^\mathrm{out}\}$, in Eq. (\ref{enhar})
are obtained then as the solutions to a single-particle Hamiltonian,
\begin{equation}
\widehat{H}_{ETF} = - \frac{\hbar^2}{2m} \nabla^2 + V_{ETF},
\label{hetf}
\end{equation}
where $V_{ETF}$ is given by Eq. (\ref{VKS}) with $\rho_{KS} ({\bf r})$
replaced by $\rho_{ETF} ({\bf r})$.
These single-particle eigenvalues will be denoted by 
$\{ \widetilde{\varepsilon}_i \}$

As is well known, the ETF-DFT does not contain shell effects 
\cite{brac2,serr,memb}. Consequently,
the corresponding density $\rho_{ETF}$ can be taken \`{a} la Strutinsky as the
smooth part, $\widetilde{\rho}$, of the KS density, $\rho_{KS}$. Accordingly,
$E_{ETF}$ is identified with the smooth part $\widetilde{E}$ in Eq.\
(\ref{enshsm}) (in the following, the ``ETF" subscript and ``~$\widetilde{~}$~"
can be used interchangeably). Since, as aforementioned,
$E_\mathrm{Harris}[\rho_{ETF}]$ approximates well [i.e., to second order in
$(\rho_{ETF}-\rho_{KS})$] the self-consistent total energy $E_{KS}[\rho_{KS}]$,
it follows from Eq.\ (\ref{enshsm}), with $E_\mathrm{Harris}[\rho_{ETF}]$ 
taken as the expression for $E_\mathrm{total}$, that the shell-correction, 
$\Delta E_{sh}$, is given by
\begin{equation}
\Delta E_{sh} = E_\mathrm{Harris}[\rho_{ETF}] - E_{ETF}[\rho_{ETF}] \equiv
E_{sh}[\widetilde{\rho}] - \widetilde{E}[\widetilde{\rho}].
\label{denhar}
\end{equation}

Defining,
\begin{equation}
T_{sh} = \sum_{i=1}^\mathrm{occ} \widetilde{\varepsilon}_i - 
\int \! \rho_{ETF} ({\bf r}) V_{ETF}({\bf r}) d{\bf r},
\label{tsh}
\end{equation}
and denoting the total energy $E_\mathrm{Harris}$ by $E_{sh}$, i.e., by identifying
\begin{equation}
E_{sh} \equiv E_\mathrm{Harris},
\label{esheqeharr}
\end{equation}
we obtain
\begin{equation}
E_{sh}[\widetilde{\rho}] = \{ T_{sh} - \widetilde{T}[\widetilde{\rho}]\}
+\widetilde{E}[\widetilde{\rho}],
\label{ensh}
\end{equation}
where $\widetilde{T}[\widetilde{\rho}]$ is the ETF kinetic energy, given to
fourth-order gradients by the expression \cite{hodg},
\begin{eqnarray}
T_{ETF}[\rho] & = & \frac{\hbar^2}{2m} \int \! 
\left\{
\frac{3}{5} (3\pi^2)^{2/3} \rho^{5/3} + 
\frac{1}{36} \frac{(\nabla \rho)^2}{\rho} +
\frac{1}{270} (3 \pi^2)^{-2/3} \rho^{1/3}  
\right. \nonumber \\
~~ & ~~ & \times 
\left. \left[ 
\frac{1}{3} \left( \frac{\nabla \rho} {\rho}
\right)^4 - \frac{9}{8} \left( \frac{\nabla \rho} {\rho} \right)^2 
\frac{\Delta \rho} {\rho} +
 \left( \frac{\Delta \rho} {\rho} \right)^2 
\right] \right\}
d {\bf r},
\label{t4th}
\end{eqnarray}
which as noted before does not contain shell effects.
Therefore, the shell correction term in Eq.\ (\ref{enshsm}) [or Eq.\
(\ref{denhar})] is given by a difference between kinetic energy terms,
\begin{equation}
\Delta E_{sh} = T_{sh}-\widetilde{T}[\widetilde{\rho}].
\label{densh}
\end{equation}

One should note that the above derivation of the shell correction does not
involve a Strutinsky averaging procedure of the kinetic energy operator.
Rather it is based on using ETF quantities as the smooth part for the density,
$\widetilde{\rho}$, and energy, $\widetilde{E}$. Other descriptions of the
smooth part may result in different shell-correction terms.

To check the accuracy of this procedure, we have compared results of 
calculations using the functional $E_{{sh}}$ [Eq.\ (\ref{ensh})]
with available Kohn-Sham calculations. In general, the optimized density from the 
minimization of the ETF-DFT functional can be obtained numerically as a solution
of the differential equation
\begin{equation}
\frac{\delta T_{ETF}[\rho]}{\delta{\rho}({\bf r})} + V_{ETF}[\rho({\bf r})]=\mu,
\label{rhoalt}
\end{equation}
where $\mu$ is the chemical potential.
As mentioned already, for the jellium DFT-SCM calculations, we often use a trial
density profile in the ETF-DFT variation which is chosen as,
\begin{equation}
\rho(r) = \frac{\rho_0}{ \left[ 1+
\exp \left(\frac{r-r_0}{\alpha}\right) \right]^{\gamma}},
\label{rho}
\end{equation}
with $r_0$, $\alpha$, and $\gamma$ as variational parameters that minimize the
ETF-DFT functional (for other closely related parametrizations, cf.\ 
Refs.\ \cite{serr,memb}). 

\begin{figure}[t]
\centering\includegraphics[width=8.cm]{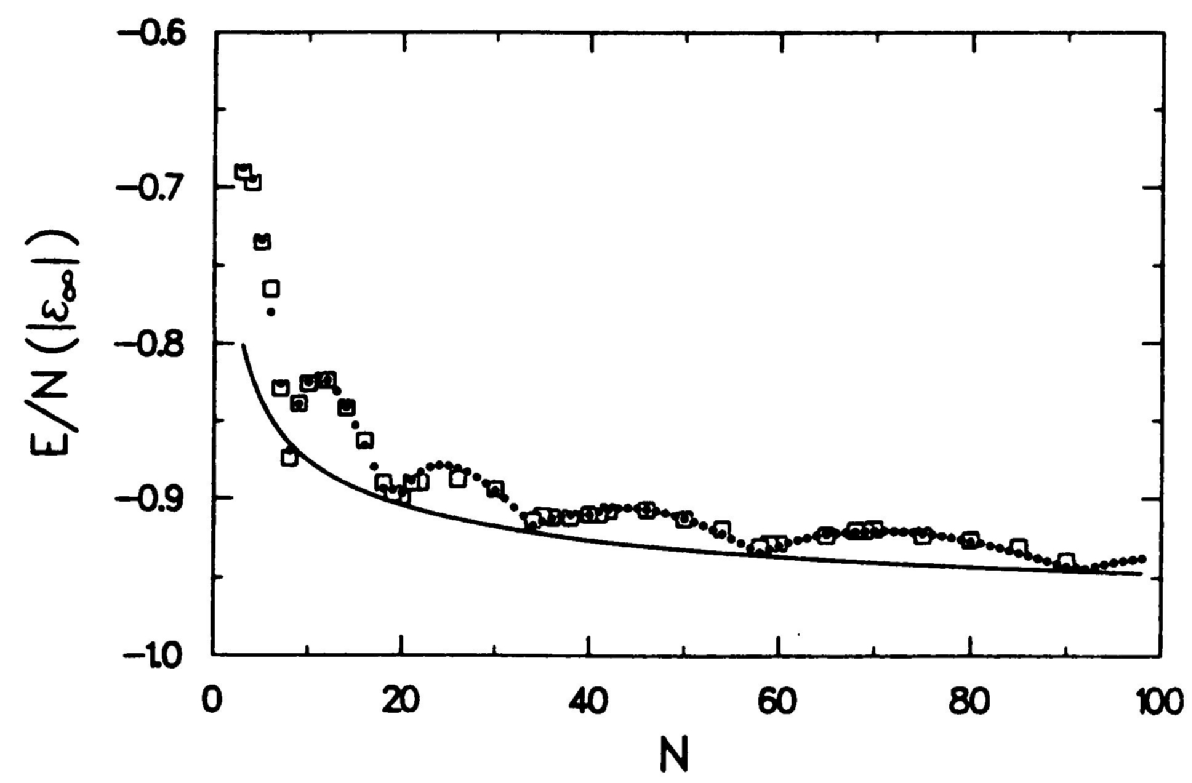}
\caption{
Total energy per atom of neutral sodium clusters (in units of the absolute 
value of the energy per atom in the bulk, $|\varepsilon_\infty|=2.252\; eV$). 
Solid circles: DFT-SCM results (see text for details). 
The solid line is the ETF result (smooth contribution).
In both cases, a spherical jellium background was used. 
Open squares: Kohn-Sham DFT results from Ref.\protect\ \protect\cite{ekar2}.
The excellent agreement (a discrepancy of only 1\%) between the
DFT-SCM and the Kohn-Sham DFT approach is to be stressed.
}
\label{fig2}
\end{figure}
\begin{figure}[t]
\centering\includegraphics[width=8.cm]{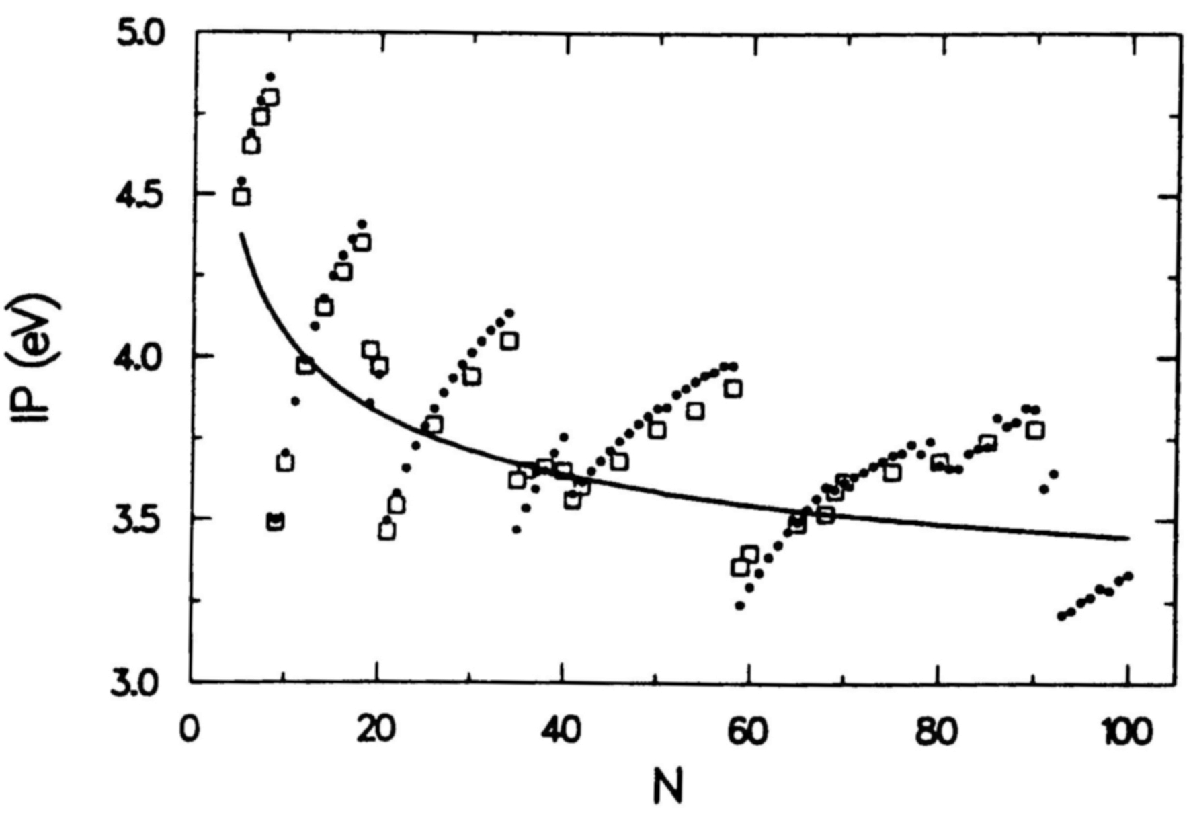}
\caption{
Ionization potentials for sodium clusters. Solid circles: IPs
calculated with the DFT-SCM (see text for details). The solid line corresponds 
to the ETF results (smooth contribution). In both cases, a spherical jellium
background was used. Open squares: Kohn-Sham DFT results from 
Ref.\protect\ \protect\cite{ekar2}. The excellent agreement 
(a discrepancy of only 1-2\%) between the DFT shell correction method and the 
full Kohn-Sham approach should be noted.
}
\label{fig3}
\end{figure}

\Fref{fig2} displays results of the present shell 
correction approach for the total energies of neutral sodium clusters. 
The results of the shell correction method for ionization potentials
of sodium clusters are displayed in \fref{fig3}. The excellent agreement 
between the oscillating results obtained via our DFT-SCM theory and the Kohn-Sham 
results (cf., e.g., Ref.\ \cite{ekar2}) is evident. To further illustrate 
the two components (smooth contribution and shell correction) entering
into our approach, we also display the smooth parts resulting from the
ETF method. (In all calculations, the Gunnarsson-Lundqvist exchange and correlation 
energy functionals were used; see Refs.\ \cite{yl1,yl2}.)

\section{Applications of DFT-SCM}
\label{appl}

\subsection{Metal clusters}
\label{metclu}

\subsubsection{Charging of metal clusters}

Investigations of metal clusters based on
DFT methods and self-consistent solutions of the Kohn-Sham 
equations (employing either a positive jellium background or
maintaining the discrete ionic cores) have contributed
significantly to our understanding of these systems \cite{dehe,barn,ekar2,andr}. 
However, even for singly negatively charged metal
clusters ($M_N^-$), difficulties may arise due to the failure of the solutions
of the KS equations to converge, since the eigenvalue of the excess
electron may iterate to a positive energy \cite{perd}. While such difficulties
are alleviated for $M_N^-$ clusters via self-interaction corrections (SIC)
\cite{alon,ekar8}, the treatment of multiply charged clusters $(M_N^{Z-},\;
Z > 1)$ would face similar difficulties in the metastability region
against electronic autodetachment through a Coulombic barrier.
In the following we are applying our DFT-SCM approach, described in the previous
section, to these systems \cite{yl1,yl2,yl3}
(for the jellium background, we assume spherical symmetry, unless otherwise stated;
for a discussion of cluster deformations, see \sref{a1}).

\subsubsection{Electron affinities and borders of stability}

The smooth multiple electron affinities $\widetilde{A}_Z$  prior to shell 
corrections are defined as the difference
in the total energies of the clusters 
\begin{equation}
\widetilde{A}_Z = \widetilde{E}(vN,vN+Z-1) - \widetilde{E}(vN,vN+Z),
\label{azsm}
\end{equation}
where $N$ is the number of atoms, $v$ is the valency
and $Z$ is the number of excess electrons in the cluster
(e.g., first and second affinities correspond to $Z=1$ and $Z=2$,
respectively). $vN$ is the total charge of the positive background. 
Applying the shell correction in Eq.\ (\ref{densh}), we calculate the full electron 
affinity as
\begin{equation}
A^{sh}_Z-\widetilde{A}_Z = \Delta E_{{sh}}(vN,vN+Z-1)-
\Delta E_{{sh}}(vN,vN+Z).
\label{azsh}
\end{equation}

\begin{figure}[t]
\centering\includegraphics[width=7.cm]{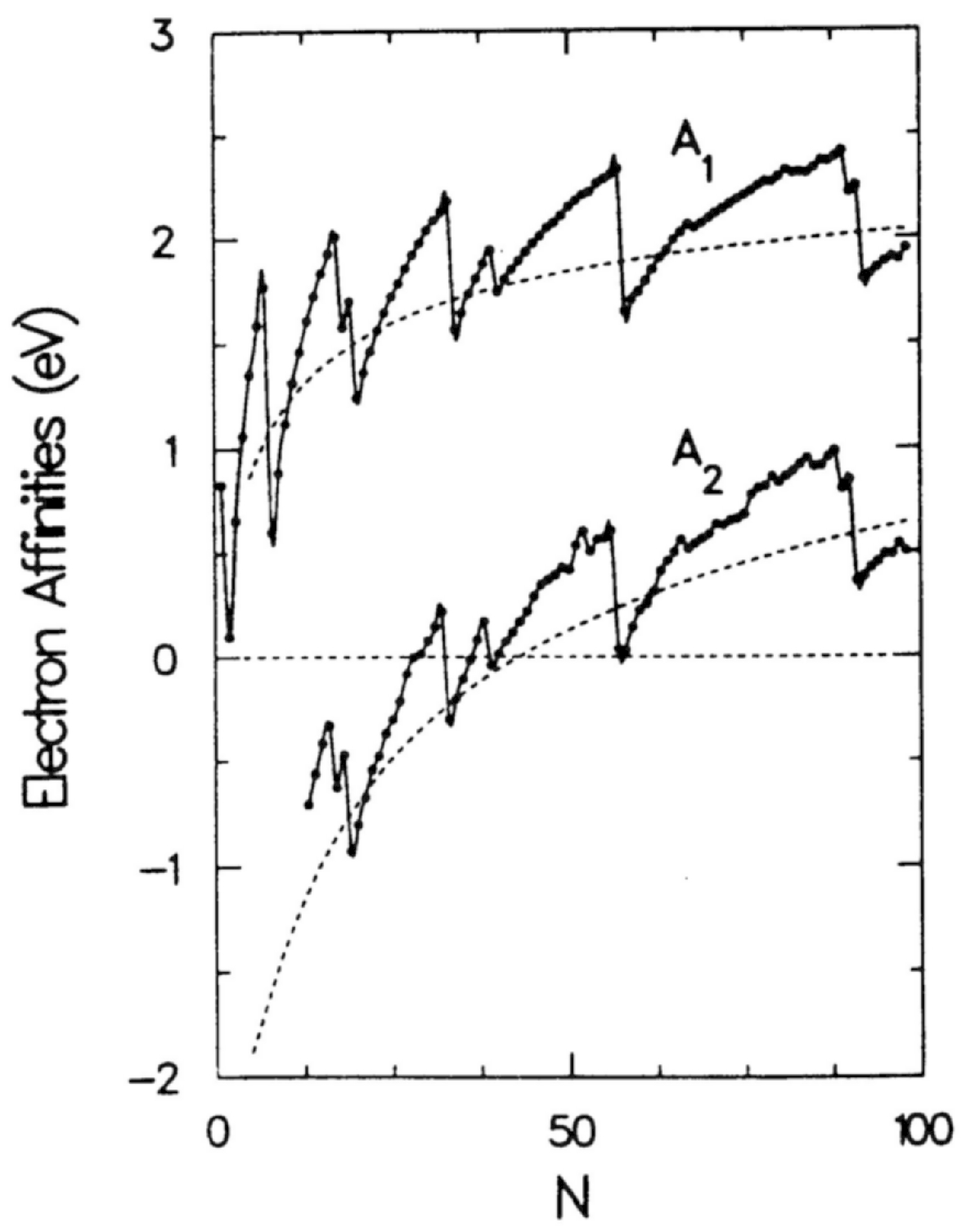}
\caption{
Calculated first ($A_1$) and second ($A_2$) electron affinities of sodium 
clusters as a function of the number of atoms $N$. Both their smooth
part (dashed lines) and the shell-corrected affinities (solid circles) 
are shown. A spherical jellium background was used.
}
\label{fig4}
\end{figure}
\begin{figure}[t]
\centering\includegraphics[width=8.cm]{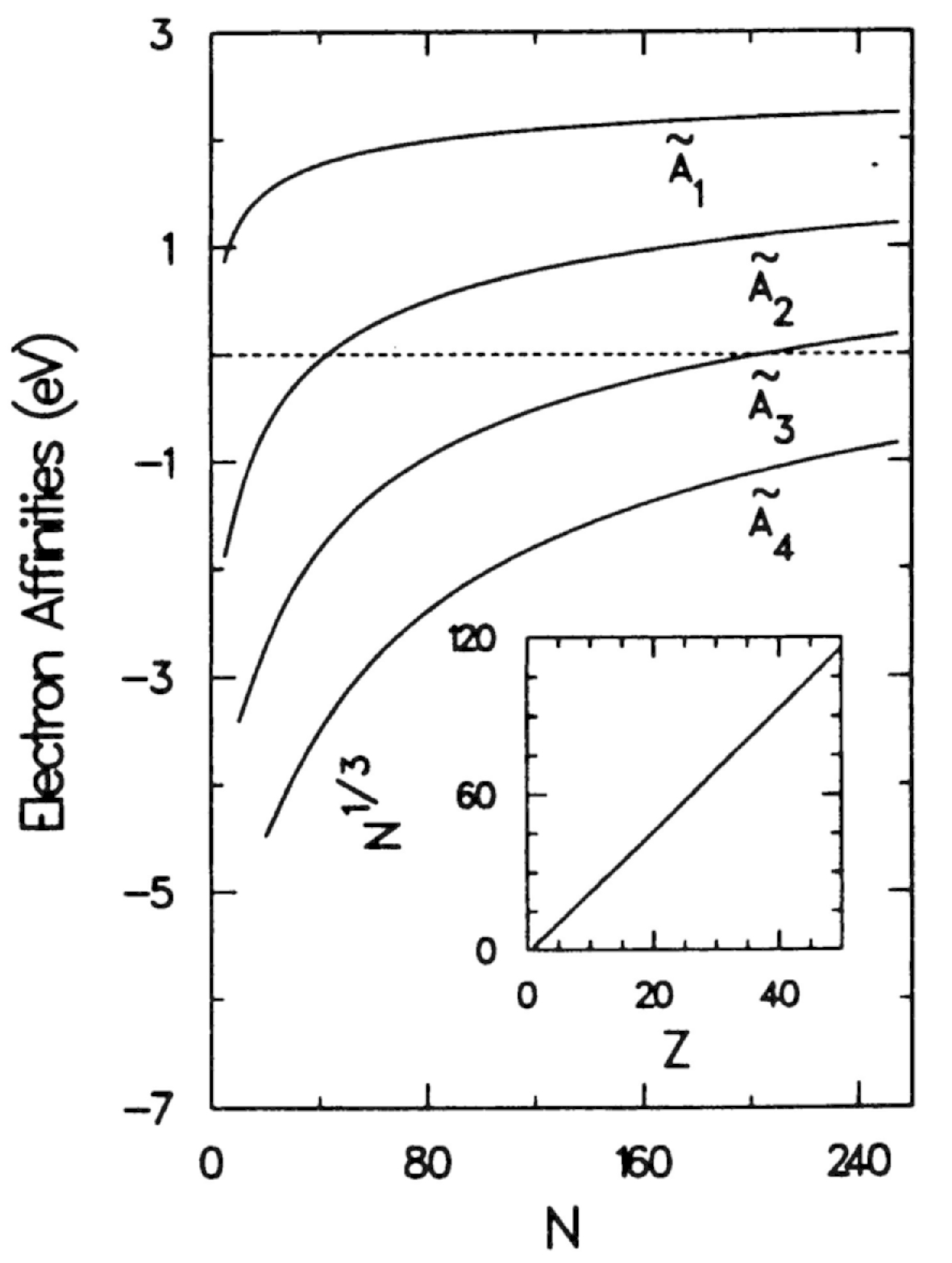}
\caption{
Calculated smooth electron affinities $\widetilde{A}_Z$, $Z=1$-4, for
sodium clusters as a function of the number of atoms $N$ ($Z$ is the
number of excess electrons). A spherical jellium background was used.
Inset: The electron {\it drip\/} line
for sodium clusters. Clusters stable against spontaneous electron
emission are located above this line.
While for spherical symmetry, as seen from \fref{fig4}, 
shell effects influence the border of stability, shell-corrected
calculations \cite{yl4} including deformations (see the appendix) yield
values close to the drip line (shown in the inset) which
was obtained from the smooth contributions.
}
\label{fig5}
\end{figure}

A positive value of the electron affinity indicates stability upon attachment
of an extra electron. \Fref{fig4} displays the smooth, as well as the shell corrected,
first and second electron affinities for sodium clusters with $N<100$. Note
that $\widetilde{A}_2$ becomes positive above a certain critical size, implying
that the second electron in doubly negatively charged sodium clusters 
with $N<N_{\rm cr}^{(2)}=43$ might not be stably attached. The shell effects, 
however, create two islands of stability about the magic clusters Na$_{32}^{2-}$ and 
Na$_{38}^{2-}$ (see $A^{sh}_2$ in \fref{fig4}). To predict the critical cluster size 
$N_{\rm cr}^{(Z)}$, which allows stable attachment
of $Z$ excess electrons, we calculated the smooth electron affinities of 
sodium clusters up to $N=255$ for $1 \leq Z \leq 4$, and display the results in 
\fref{fig5}. We observe that $N_{\rm cr}^{(3)}=205$, while $N_{\rm cr}^{(4)} > 255$. 

The similarity of the shapes of the curves in \fref{fig5}, and the regularity
of distances between them, suggest that the smooth 
electron affinities can be fitted by a general expression of the form:
\begin{equation}
\widetilde{A}_{Z} = \widetilde{A}_1 - \frac{(Z-1)e^2}{R+\delta} 
           = W - \beta \frac{e^2}{R+\delta} - \frac{(Z-1)e^2}{R+\delta},
\label{asman}
\end{equation}
where the radius of the positive background is $R=r_sN^{1/3}$.
From our fit, we find that the constant $W$ corresponds to the bulk work function. 
In all cases, we find $\beta=5/8$, which suggests a close analogy with
the classical model of the image charge \cite{wood,stav}. For
the spill-out parameter, we find a weak size dependence as $\delta =
\delta_0 + \delta_2/R^2$. The contribution of 
$\delta_2/R^2$, which depends on $Z$, is of importance only for
smaller sizes and does not affect substantially the 
critical sizes (where the curve crosses the zero line), and consequently
$\delta_2$ can be neglected in such estimations. Using the
values obtained by us for $\widetilde{A}_1$ of sodium clusters (namely,
$W=2.9 \;eV$ which is also the value obtained by KS-DFT calculations for 
an infinite planar surface \cite{perd2}, 
$\delta_0=1.16 \; a.u.$; with $ R=r_s N^{1/3}$,
and $r_s=4.00\; a.u.$), we find
for the critical sizes when the l.h.s. of Eq.\ (\ref{asman}) is set equal
to zero, $N_{\rm cr}^{(2)}=44$, $N_{\rm cr}^{(3})=202$, 
$N_{\rm cr}^{(4)}=554$, and $N_{\rm cr}^{(5)}=1177$, in very good 
agreement with the values obtained directly from \fref{fig5}.

The curve that specifies $N^{(Z)}_{\rm cr}$ in the $(Z,N)$ plane defines
the border of stability for spontaneous electron decay. 
In nuclei, such borders of stability against spontaneous proton or
neutron emission are known as nucleon drip lines \cite{siem}. For the case of 
sodium clusters, the electron drip line is displayed in the inset of \fref{fig5}.

\subsubsection{Critical sizes for potassium and aluminum}

While in this investigation we have used sodium clusters as a test system,
the methodology and conclusions extend to other materials as well.
Thus given a calculated or measured bulk work function $W$, and a spill-out
parameter ($\delta_0$ typically of the order of 1-2 $a.u.$, and neglecting
$\delta_2$), one can use Eq.\ (\ref{asman}), with $\widetilde{A}_Z=0$, to predict
critical sizes for other materials. For example, our calculations for 
potassium ($r_s=4.86\; a.u.$) give fitted values $W=2.6 \; eV$ (compared
to a KS-DFT value of 2.54 $eV$ for a semi-infinite planar surface with
$r_s=5.0\; a.u.$ \cite{perd2}) and $\delta_0=1.51\; a.u.$ for $\delta_2=0$, 
yielding $N_{\rm cr}^{(2)}=33$, $N_{\rm cr}^{(3)}=152$, and 
$N_{\rm cr}^{(4)}=421$.

As a further example, we give our results for a trivalent metal, i.e.\ 
aluminum ($r_s=2.07\;a.u.$), for which our fitted values are $W=3.65\; eV$
(compared to a KS-DFT value of 3.78 $eV$ for a semi-infinite
plane surface, with $r_s=2.0$ $a.u.$ \cite{perd2}) and
$\delta_0=1.86\; a.u.$ for $\delta_2=0$,
yielding $N_{\rm cr}^{(2)}=40$ (121 electrons),
$N_{\rm cr}^{(3)}=208$ (626 electrons), and $N_{\rm cr}^{(4)}=599$
(1796 electrons).

\subsubsection{Metastability against electron autodetachment}

\begin{figure}[t]
\centering\includegraphics[width=7.cm]{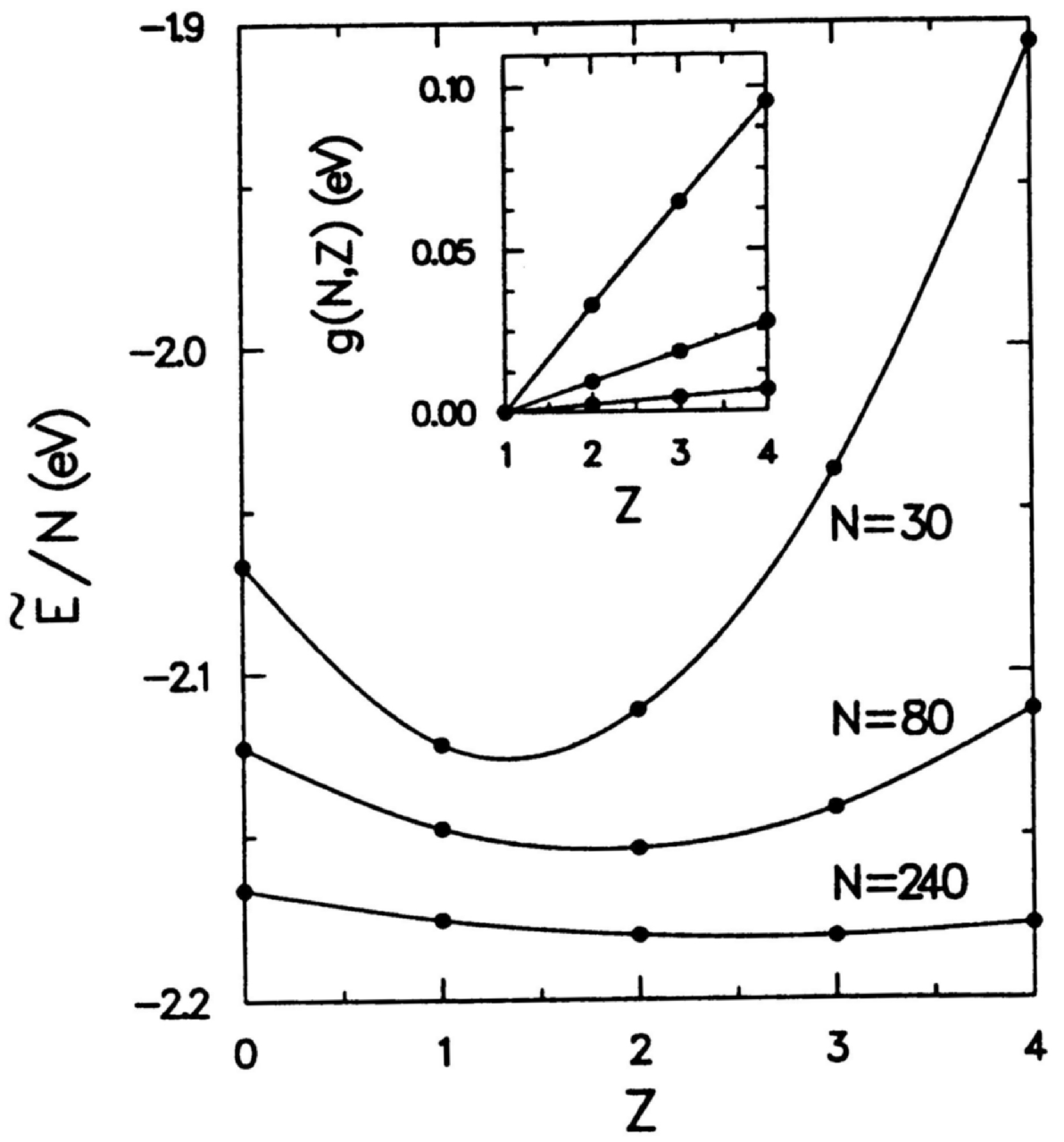}
\caption{
Calculated smooth total energy per atom as a function of the excess negative
charge $Z$ for the three families of sodium clusters with $N=30$, $N=80$, and
$N=240$ atoms. A spherical jellium background was used.
As the straight lines in the inset demonstrate, the curves are 
parabolic. We find that they can be fitted by Eq. (\ref{etz}). See text for an 
explanation of how the function $g(N,Z)$ was extracted from the calculations.
}
\label{fig6}
\end{figure}

The multiply charged anions with negative affinities do not necessarily exhibit 
a positive total energy. To illustrate this point, we display in \fref{fig6}
the calculated total energies per atom ($\widetilde{E}(N,Z)/N$)
as a function of excess charge ($Z$) for clusters containing 30, 80, and 
240 sodium atoms. These sizes allow for exothermic 
attachment of maximum one, two, or three excess electrons, respectively.

As was the case with the electron affinities, the total-energy curves 
in \fref{fig6} show a remarkable regularity, suggesting a
parabolic dependence on the excess charge. To test this conjecture, we have
extracted from the calculated total energies the quantity
$g(N,Z)=G(N,Z)/N$ where 
$G(N,Z)=[\widetilde{E}(N,Z)-\widetilde{E}(N,0)]/Z + \widetilde{A}_1(N)$, 
and have plotted it in the inset of \fref{fig6} as a function
of the excess negative charge $Z$. The dependence is linear to a remarkable
extent; for $Z=1$ all three lines cross the energy axis at zero. 
Combined with the results on the electron affinities, this
indicates that the total energies have the following dependence
on the excess number of electrons ($Z$):
\begin{equation}
\widetilde{E}(Z)=\widetilde{E}(0)-\widetilde{A}_1 Z +
\frac{Z(Z-1)e^2}{2(R+\delta)},
\label{etz}
\end{equation}
where the dependence on the number of atoms in the cluster is not
explicitly indicated.

This result is remarkable in its analogy with the classical image-charge
result of van Staveren {\it et al.\/} \cite{stav}.
Indeed, the only difference amounts to the spill-out parameter 
$\delta_0$ and to the weak dependence on $Z$ through $\delta_2$. 
This additional $Z$-dependence becomes negligible already for the case of 
30 sodium atoms.
 
\begin{figure}[t]
\centering\includegraphics[width=7.cm]{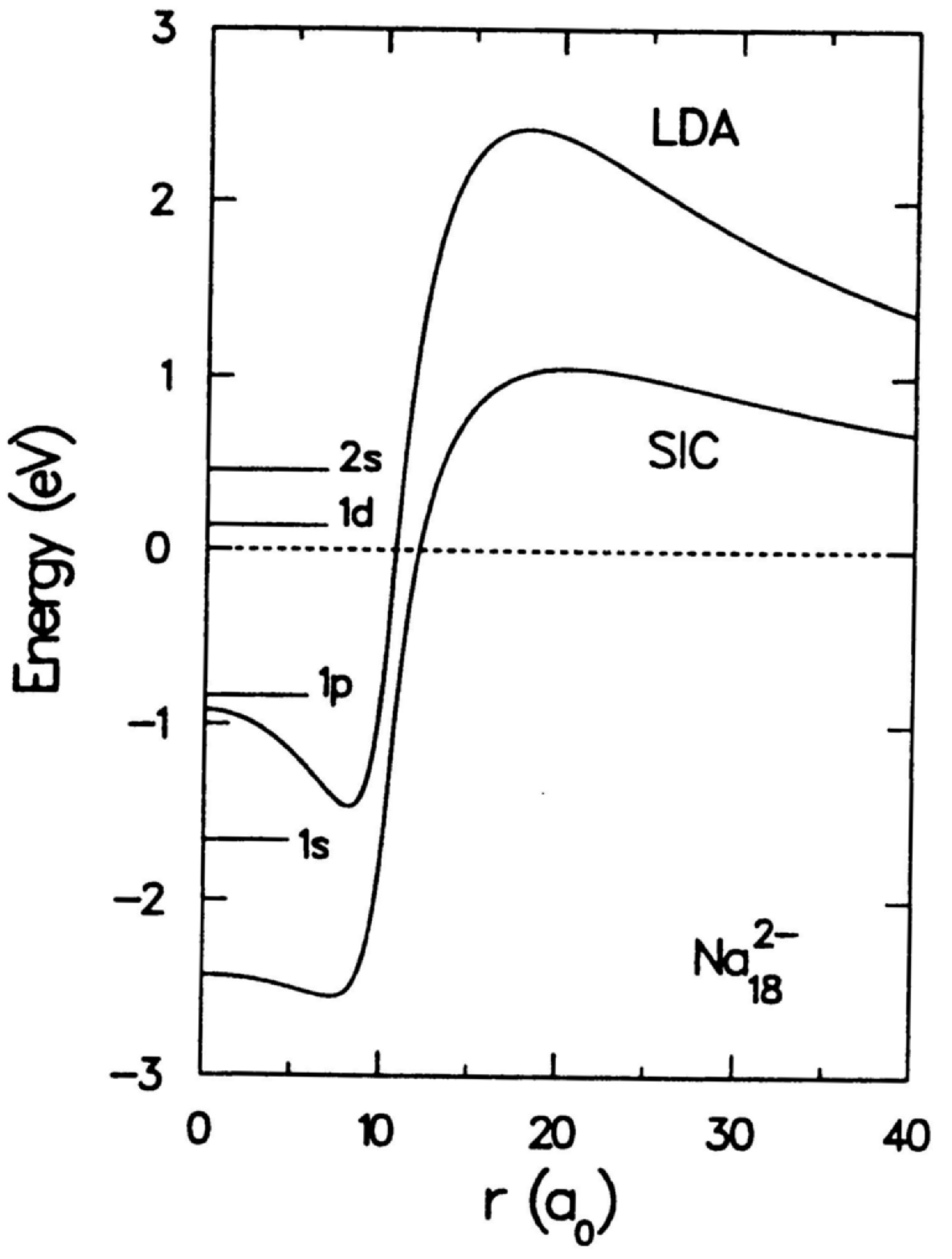}
\caption{
The DFT (LDA) and the corresponding self-interaction corrected potential for 
the metastable Na$_{18}^{2-}$ cluster. A spherical jellium background was used.
The single-particle levels of the SIC 
potential are also shown. Unlike the LDA, this latter potential exhibits the 
correct asymptotic behavior. The $2s$ and $1d$ electrons can be
emitted spontaneously by tunneling through the Coulombic barrier of the SIC 
potential. Distances in units of the Bohr radius, $a_0$.
The specified single-particle levels are associated with the SIC potential.
}
\label{fig7}
\end{figure}

For metastable multiply-charged cluster anions, electron emission
(autodetachment) will occur via tunneling through a barrier (shown in
\fref{fig7}). However, to reliably estimate the electron emission, it is 
necessary to correct the LDA effective potential for self-interaction effects.
We performed a self-interaction correction of the Amaldi type \cite{perd} for the 
Hartree term and extended it to the exchange-correlation contribution to the total
energy as follows: $E_{\rm xc}^{\rm SIC}[\rho]=
E_{\rm xc}^{\rm LDA}[\rho]-N_e E_{\rm xc}^{\rm LDA}[\rho/N_e]$,
where $N_e=vN+Z$ is the total number of electrons. 
This self-interaction correction is akin
to the orbitally-averaged-potential method \cite{perd}. Minimizing
the SIC energy functional for the parameters $r_0$, $\alpha$, and
$\gamma$, we obtained the effective SIC potential
for Na$_{18}^{2-}$ shown in \fref{fig7}, which exhibits the physically correct
asymptotic behavior \cite{note}.

The spontaneous electron emission through the Coulombic barrier is analogous
to that occurring in proton radioactivity from neutron-deficient nuclei
\cite{hofm}, as well as in alpha-particle decay. The transition rate is
$\lambda=\ln 2 / T_{1/2} = \nu P $, where $\nu$ is the attempt frequency
and $P$ is the transmission coefficient calculated in the WKB method 
(for details, cf. Ref.\ \cite{hofm}). For the $2s$ electron in Na$_{18}^{2-}$
(cf. \fref{fig7}), we find $\nu = 0.73\;10^{15}\; Hz$ and $P=4.36\; 10^{-6}$,
yielding $T_{1/2}=2.18\; 10^{-10}\; s$. For a cluster size closer to the drip 
line (see \fref{fig5}), e.g. Na$_{35}^{2-}$, we find $T_{1/2} = 1.13 \; s$.

Finally, the exression in Eq.\ (\ref{etz}) for the total energy can be naturally 
extended to the case of multiply {\it positively\/} charged metal clusters by 
setting $Z=-z$, with $z>0$. The ensuing equation retains the same dependence
on the excess positive charge $z$, 
but with the negative value of the first affinity, $-\widetilde{A}_1$,
replaced by the positive value of the first ionization potential,
$\widetilde{I}_1=W+(3/8)e^2/(R+\delta)$, a result that has been 
suggested from earlier measurements on multiply charged potassium
cations \cite{brec}. Naturally, the spill-out parameter $\delta$
assumes different values than in the case of the anionic clusters.

\subsection{Neutral and multiply charged fullerenes} 
\label{full}

\subsubsection{Stabilized jellium approximation - The generalized DFT-SCM}

Fullerenes and related carbon structures have been extensively investigated 
using {\it ab initio\/} density-functional-theory methods and self-consistent 
solutions of the Kohn-Sham (KS) equations \cite{trou,koha}. For metal clusters, 
replacing the ionic cores with a uniform jellium background was found to
describe well their properties within the KS-DFT method \cite{dehe2}. 
Motivated by these results, several attempts
to apply the jellium model in conjunction with DFT to investigations
of fullerenes have appeared recently \cite{yaba,lipp,pusk,yl7}.
Our approach \cite{yl7} differs from the earlier ones in several aspects and,
in particular, in the adaptation to the case of finite systems 
of the stabilized-jellium (or structureless pseudopotential)
energy density functional (see Eq.\ (\ref{epseu}) below and
Ref.\ \cite{perd}).

An important shortcoming of the standard jellium approximation for
fullerenes (and other systems with high density, i.e., small
$r_s$) results from a well-known property of the jellium at
high electronic densities, namely that the jellium is unstable and 
yields negative surface-energy contribution to the total energy
\cite{perd}, as well as unreliable values for the total energy.
These inadequacies of the standard jellium model can be rectified by
pseudopotential corrections. A modified-jellium approach
which incorporates such pseudopotential corrections
and is particularly suited for our purposes here, is the 
{\it structureless pseudopotential\/} model or
{\it stabilized jellium\/} approximation developed in Ref.\ \cite{perd}.

In the stabilized jellium, the total energy $E_{pseudo}$, as a
functional of the electron density $\rho({\bf r})$, is given
by the expression
\begin{equation}
E_{pseudo}[\rho,\rho_+] = E_{jell}[\rho,\rho_+] +
\langle \delta \upsilon \rangle_{WS}
\int \rho({\bf r}) {\cal U}({\bf r}) d{\bf r}
-\widetilde{\varepsilon} \int \rho_+({\bf r}) d{\bf r},
\label{epseu}
\end{equation}
where by definition the function ${\cal U}({\bf r})$ equals unity inside, 
but vanishes, outside the jellium volume. 
$\rho_+$ is the density of the positive jellium background
(which for the case of C$_{60}$ is taken as a spherical shell,
of a certain width $2d$, centered at 6.7 $a.u.$ ).
$E_{pseudo}$ in Eq.\ (\ref{epseu}) is the standard jellium-model total energy, 
$E_{jell}$, modified by two corrections. The first correction adds the effect
of an average (i.e., averaged over the volume of a Wigner-Seitz cell) 
difference potential, $\langle \delta \upsilon \rangle_{WS} \cal{U} ({\bf r})$, 
which acts on the electrons in addition to the standard jellium attraction
and is due to the atomic pseudopotentials (in this work, we use the
Ashcroft empty-core pseudopotential, specified by a core radius
$r_c$, as in Ref.\ \cite{perd}).
The second correction subtracts 
from the jellium energy functional the spurious electrostatic
self-repulsion of the positive background within each cell; this term makes
no contribution to the effective electronic potential.

Following Ref.\ \cite{perd}, the bulk stability condition {\bf (}Eq.\ (25) in
Ref.\ \cite{perd}{\bf )} determines the value of the pseudopotential core
radius $r_c$, as a function of the bulk Wigner-Seitz radius $r_s$.
Consequently, the difference potential can be expressed
solely as a function of $r_s$ as follows (energies in $Ry$, 
distances in $a.u.$):
\begin{equation}
\langle \delta \upsilon \rangle_{WS} = 
-\frac{2}{5} \left ( \frac{9\pi}{4}\right )^{2/3} r_s^{-2} + 
\frac{1}{2\pi} \left ( \frac{9\pi}{4} \right )^{1/3} r_s^{-1} +
\frac{1}{3} r_s \frac{d\varepsilon_c}{dr_s},
\label{dvws}
\end{equation}
where $\varepsilon_c$ is the per particle electron-gas correlation energy
(in our calculation, we use the Gunnarsson-Lundqvist exchange and correlation 
energy functionals; see Refs.\ \cite{yl1,yl2}).

The electrostatic self-energy, $\widetilde{\varepsilon}$, per unit charge 
of the uniform positive jellium is given by
\begin{equation}
\widetilde{\varepsilon} = 6\upsilon^{2/3}/5r_s,
\label{vare}
\end{equation}
where $\upsilon$ is the valence of the atoms ($\upsilon=4$ for carbon).

\subsubsection{ETF electron-density profile}

To apply the ETF-DFT method to carbon fullerenes, we generalize it by 
employing potential terms according to the stabilized-jellium functional 
in Eq.\ (\ref{epseu}).

\begin{figure}[t]
\centering\includegraphics[width=7.cm]{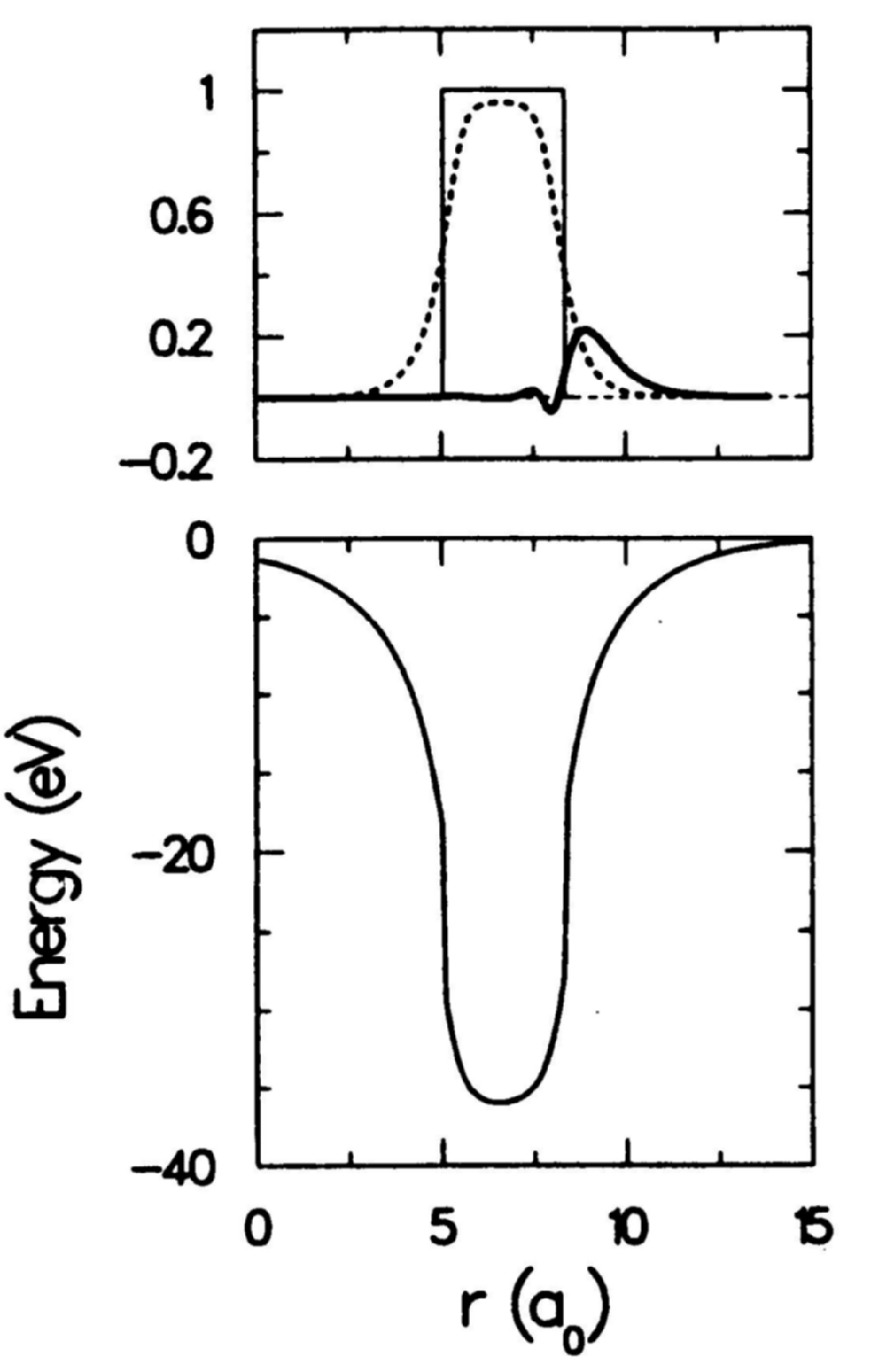}
\caption{
Bottom panel: The stabilized-jellium LDA potential obtained by the ETF
method for the neutral C$_{60}$ molecule. The Wigner-Seitz radius for
the jellium bacground is 1.23 $a.u.$ Note the asymmetry of the potential
about the minimum. The associated difference potential
$\protect \langle \delta \upsilon \protect\rangle_{WS}=-9.61 \; eV$.
\protect\\
Top panel: Solid line: Radial density of the positive jellium background.
Dashed line: ETF electronic density. Note its asymmetry about the maximum.
Thick solid line: The difference (multiplied by 10) of electronic ETF densities
between C$_{60}^{5-}$ and C$_{60}$. It illustrates that the excess
charge accumulates in the outer perimeter of the total electronic
density. All densities are normalized to the density of the positive
jellium background.
}
\label{fig8}
\end{figure}

Another required generalization consists in employing
a parametrized electron-density profile that accounts for the hollow
cage-structure of the fullerenes. 
Such a density profile is provided by the following adaptation of a 
generalization of an inverse Thomas-Fermi distribution,
used earlier in the context of nuclear physics \cite{gram}, i.e.,
\begin{equation}
\rho (r) = \rho_0 
\left( 
\frac{ F_{i,o} \sinh [w_{i,o}/\alpha_{i,o}]}
{\cosh [w_{i,o}/\alpha_{i,o}] + \cosh [(r-R)/\alpha_{i,o}]} 
\right)^{\gamma_{i,o}},
\label{rgram}
\end{equation}
where $R=6.7\;a.u.$ is the radius of the fullerene cage.
$w$, $\alpha$, and $\gamma$ are variables to be determined
by the ETF-DFT minimization. For $R=0$ and large values of $w/\alpha$,
expression (\ref{rgram}) approaches the more familiar 
inverse Thomas-Fermi distribution, with
$w$ the width, $\alpha$ the diffuseness and $\gamma$
the asymmetry of the profile around $r=w$.
There are a total of six parameters to be determined, since
the indices $(i,o)$ stand for the regions inside ($r < R$) and 
outside ($r>R$) the fullerene cage. 
$F_{i,o} = (\cosh [w_{i,o}/\alpha_{i,o}]+1)/\sinh [w_{i,o}/\alpha_{i,o}]$ 
is a constant guaranteeing that the two parts of the curve join smoothly
at $r = R$.
The density profile in Eq.\ (\ref{rgram}) peaks at $r=R$ and then falls towards
smaller values both inside and outside the cage 
(see top panel of \fref{fig8}).

\subsubsection{Shell correction and icosahedral splitting}

To apply the SCM to the present case, the potential $V_{ETF}$ in 
Eq.\ (\ref{tsh}) is replaced by the stabilized-jellium LDA potential
shown in \fref{fig8}. After some rearrangements, the shell-corrected total energy 
$E_{{sh}}[\widetilde{\rho}]$ in the stabilized-jellium case 
can be written in functional form as follows
[compare to Eq.\ (\ref{ensh}), see also Eq.\ (\ref{enhar})].
\begin{eqnarray}
E_{{sh}}[\widetilde{\rho}]=
\sum_i \widetilde{\varepsilon}_i
& - & \int \! \left\{ 
\frac{1}{2} \widetilde{V}_H({\bf r}) + \widetilde{V}_{{xc}}({\bf r})
\right\}
\widetilde{\rho}({\bf r}) d\/{\bf r} \nonumber \\
~ & + &
 \int \widetilde{{\cal E}}_{{xc}} 
[\widetilde{\rho}({\bf r})] d\/{\bf r}
+ E_I - \widetilde{\varepsilon} \int \rho_+({\bf r}) d{\bf r},
\label{enshsj}
\end{eqnarray}

Heretofore, the point-group icosahedral symmetry of C$_{60}$
was not considered, since the molecule was treated as a spherically symmetric 
cage. This is a reasonable zeroth-order approximation as noticed by several
authors \cite{trou,pusk,gall,hadd}. However, considerable improvement
is achieved when the effects of the point-group icosahedral symmetry 
are considered as a next-order correction (mainly the
lifting of the angular momentum degeneracies).

The method of introducing the icosahedral splittings is that of the
crystal field theory \cite{gerl}. 
Thus, we will use the fact that the bare electrostatic
potential from the ionic cores, considered as point charges, 
acting upon an electron, obeys the
well-known expansion theorem \cite{gerl}
\begin{equation}
U({\bf r}) = - \upsilon e^2 \sum_i \frac{1}{|{\bf r}- {\bf r}_i|} =
- \sum_{l=0}^{\infty} \sum_{m=-l}^{l} \kappa_l (r) C_l^m Y_l^m (\theta,\phi),
\label{cryf}
\end{equation}
where the angular coefficients $C_l^m$ are given through the angular
coordinates $\theta_i, \phi_i$ of the carbon atomic cores, namely,
\begin{equation}
C_l^m = \sum_i Y_l^{m*} (\theta_i, \phi_i),
\label{clm}
\end{equation}
where $*$ denotes complex conjugation.

\begin{figure}[t]
\centering\includegraphics[width=8.cm]{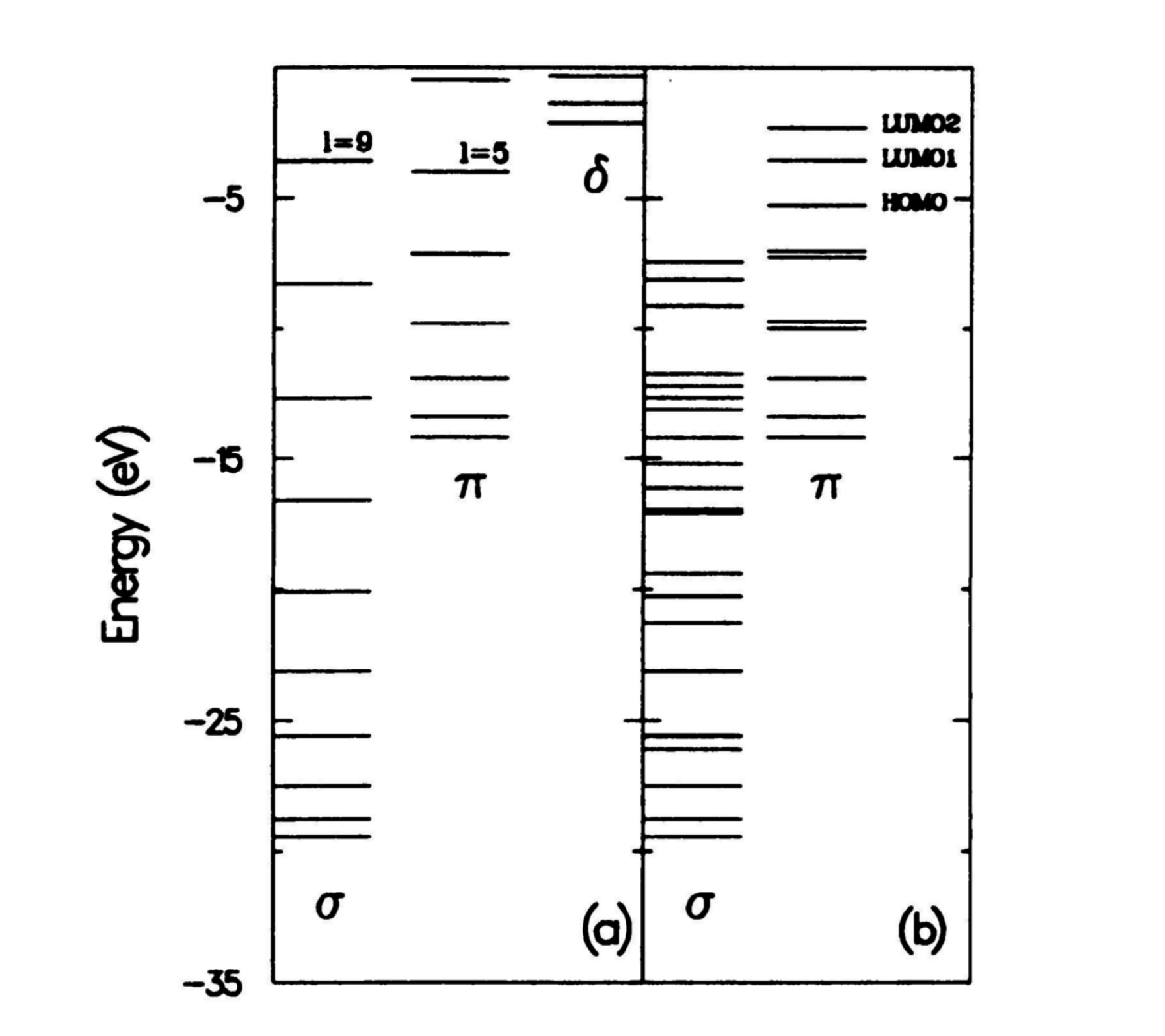}
\caption{
(a) The single-particle levels of the ETF-LDA potential for C$_{60}$ shown 
in \fref{fig8}. Because of the spherical symmetry, they are characterized
by the two principle quantum numbers $n_r$ and $l$, where $n_r$ is the
number of radial nodes and $l$ the angular momentum. They are grouped 
in three bands labeled $\sigma$ ($n_r=0$), $\pi$ ($n_r=1$), and $\delta$ 
($n_r=2$). Each band starts with an $l=0$ level. \protect\\
(b) The single-particle levels for C$_{60}$ after the icosahedral
splittings are added to the spectra of (a). The tenfold degenerate 
HOMO (h$_u$) and the sixfold degenerate LUMO1 (t$_{1u}$) and LUMO2 
(t$_{1g}$) are denoted; they originate from the spherical $l=5$ 
and $l=6$ (t$_{1g}$) $\pi$ levels displayed in panel (a).
For the $\sigma$ electrons, the icosahedral perturbation strongly splits
the $l=9$ level of panel (a). There result five sublevels which straddle the
$\sigma$-electron gap as follows: two of them (the eightfold degenerate g$_u$, 
and the tenfold degenerate h$_u$) move down and are fully occupied
resulting in a shell closure (180 $\sigma$ electrons in total). 
The remaining unoccupied levels, originating 
from the $l=9$ $\sigma$ level, are sharply shifted upwards and
acquire positive values.
}
\label{fig9}
\end{figure}

We take the radial parameters $\kappa_l(r)$ as constants, and determine
their value by adjusting the icosahedral single-particle spectra 
$\varepsilon_i^{{ico}}$ to reproduce 
the pseudopotential calculation of Ref.\ \cite{trou},
which are in good agreement with experimental data.
Our spectra without and with icosahedral splitting are shown in \fref{fig9}(a)
and \fref{fig9}(b), respectively.

We find that a close reproduction of the results of {\it ab initio\/}
DFT calculations \cite{trou,rose,ye} is achieved when 
the Wigner-Seitz radius for the jellium background is $1.23$ $a.u.$
The shell corrections, $\Delta E^{ico}_{sh}$, 
including the icosahedral splittings are 
calculated using the icosahedral single-particle energies 
$\varepsilon_i^{{ico}}$ in Eq.\ (\ref{tsh}). The average quantities
($\widetilde{\rho}$ and $\widetilde{V}$) are
maintained as those specified through the ETF variation
with the spherically symmetric profile of Eq.\ (\ref{rgram}).
This is because the first-order correction to the total energy (resulting from 
the icosahedral perturbation) vanishes, since the integral over the sphere of a 
spherical harmonic $Y_{l}^{m}\;(l>0)$ vanishes.

\subsubsection{Ionization potentials and electron affinities}

Having specified the appropriate Wigner-Seitz radius $r_s$ and 
the parameters $\kappa_l$ 
of the icosahedral crystal field through a comparison with the 
pseudopotential DFT calculations for the neutral C$_{60}$, we calculate 
the total energies of the cationic and anionic species by allowing for a 
change in the total electronic charge, namely by imposing the constraint
\begin{equation}
4 \pi \int \rho(r) r^2 dr =240 \pm x,
\label{norm}
\end{equation}
where $\rho(r)$ is given by Eq.\ (\ref{rgram}).
The shell-corrected and icosahedrally perturbed
first and higher ionization potentials $I_x^{{ico}}$
are defined as the difference of the ground-state
shell-corrected total energies $E_{{sh}}^{{ico}}$ as follows:
\begin{equation}
I_x^{{ico}} = E_{{sh}}^{{ico}}(N_e=240-x; Z=240)
-E_{{sh}}^{{ico}}(N_e=240-x+1; Z=240),
\label{ips}
\end{equation}
where $N_e$ is the number of electrons in the system and $x$ is the number 
of excess charges on the fullerenes (for the excess charge, we will 
find convenient to use two different notations $x$ and $z$ related as
$x=|z|$. A negative value of $z$ corresponds to positive excess charges).
$Z=240$ denotes the total positive charge of the jellium background.

The shell-corrected and icosahedrally perturbed first 
and higher electron affinities $A_x^{{ico}}$ are similarly defined as 
\begin{equation}
A_x^{{ico}} = E_{{sh}}^{{ico}}(N_e=240+x-1; Z=240)
-E_{{sh}}^{{ico}}(N_e=240+x; Z=240).
\label{eas}
\end{equation}

We have also calculated the corresponding average quantities
$\widetilde{I}_x$ and $\widetilde{A}_x$, which result from the ETF
variation with spherical symmetry (that is without shell and icosahedral
symmetry corrections). Their definition is the same as in
Eq.\ (\ref{ips}) and Eq.\ (\ref{eas}), but with the index $sh$ replaced by
a tilde and the removal of the index $ico$. 

In our calculations of the charged fullerene molecule, the $r_s$ value and the 
icosahedral splitting parameters {\bf (}$\kappa_l$, see Eq. (\ref{cryf}),
and discussion below it{\bf )} were taken as those which were determined
by our calculations of the neutral molecule, discussed in the previous
section. The parameters which specify the ETF electronic density 
{\bf (}Eq. (\ref{rgram}){\bf )} are optimized for the charged molecule,
thus allowing for relaxation effects due to the excess charge. This
procedure is motivated by results of previous electronic structure
calculations for C$_{60}^+$ and C$_{60}^-$ \cite{rose,ye}, 
which showed that the icosahedral 
spectrum of the neutral C$_{60}$ shifts almost rigidly upon charging
of the molecule.

Shell-corrected and ETF calculated values of ionization potentials
and electron affinities, for values of the excess charge up to 12 units, are 
summarized in \tref{tbl1} (for $r_s=1.23\; a.u.$)

\begin{table*}[t]
\tbl{ETF (spherically averaged, denoted by a tilde) and shell-corrected 
(denoted by a superscript $ico$ to indicate that the icosahedral
splittings of energy levels have been included)
IPs and EAs of fullerenes C$^{x \pm}_{60}$. Energies in $eV$. 
$r_s=1.23$ $a.u.$}
{\begin{tabular}{ccccc}
\toprule
~ & ~ & ~ & ~ & ~  \\
$x$ & $\widetilde{I}_x$ & $I^{{ico}}_x$ &
$\widetilde{A}_x$ & $ A^{{ico}}_x$ \\ \colrule 
1 &  5.00  &  7.40 & 2.05 & 2.75 \\
2 &  7.98  & 10.31 & $-$0.86 & $-$0.09 \\
3 & 10.99  & 13.28 & $-$3.75 & $-$2.92 \\
4 & 14.03  & 16.25 & $-$6.60 & $-$5.70 \\
5 & 17.09  & 19.22 & $-$9.41 & $-$8.41 \\
6 & 20.18  & 22.20 & $-$12.19 & $-$11.06 \\
7 & 23.29  & 25.24 & $-$14.94 & $-$14.85 \\
8 & 26.42  & 28.31 & $-$17.64 & $-$17.24 \\
9 & 29.57  & 31.30 & $-$20.31 & $-$19.49 \\
10 & 32.73 & 34.39 & $-$22.94 & $-$21.39 \\
11 & 35.92 & 39.36 & $-$25.53 & $-$22.93 \\
12 & 39.12 & 42.51 & $-$28.07 & $-$23.85 \\ \botrule
\end{tabular}}
\label{tbl1}
\end{table*}

\subsubsection{Charging energies and capacitance of fullerenes}

\begin{figure}[t]
\centering\includegraphics[width=7.cm]{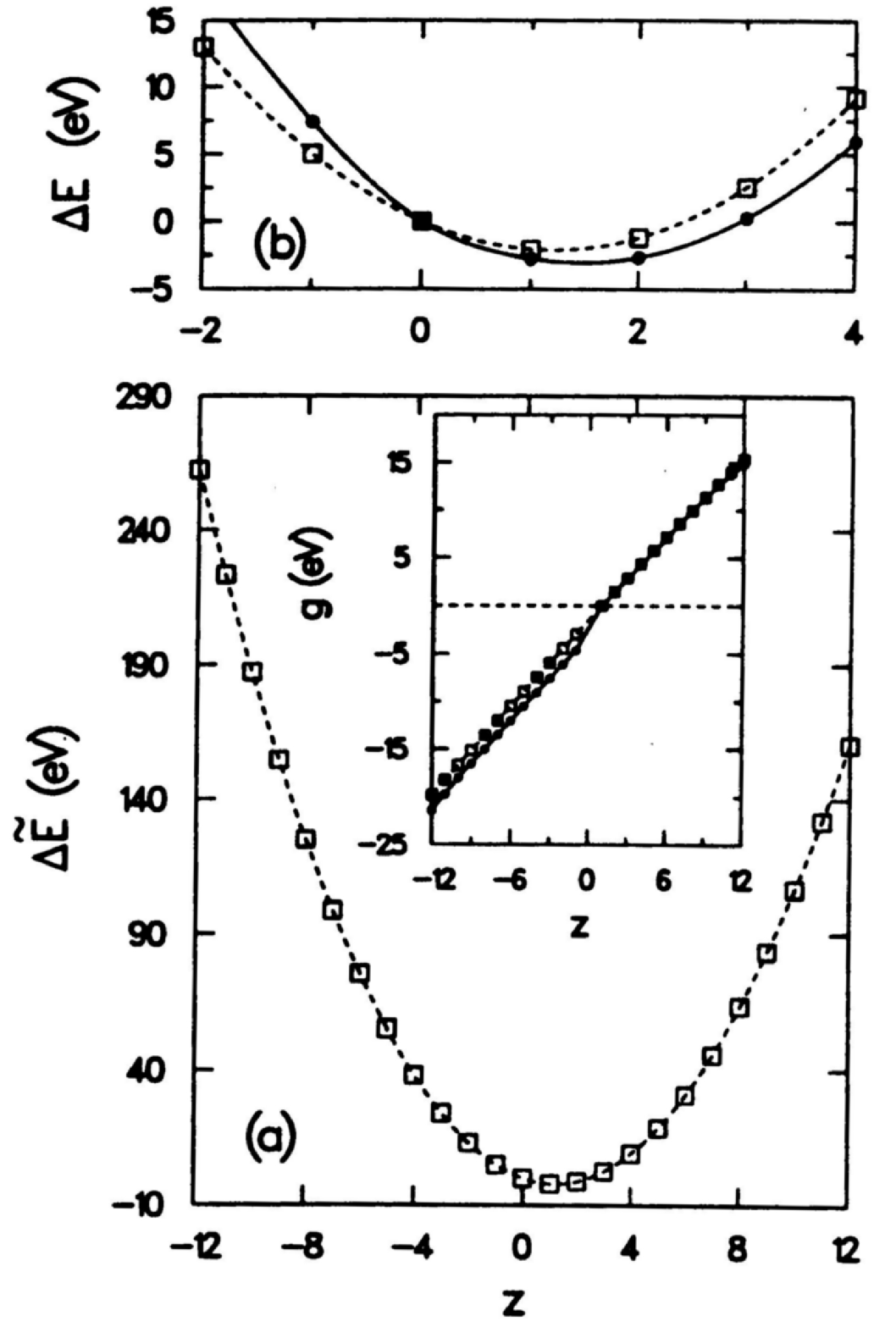}
\caption{
(a) ETF-DFT total energy differences (appearance energies)
$\Delta \widetilde{E}(z)=\widetilde{E}(z)-\widetilde{E}(0)$
as a function of the excess charge $z$ 
($z<0$ corresponds to positive excess charge). 
Inset: The ETF function $\widetilde{g}(z)$ (open squares),
and the shell-corrected 
function $g^{{ico}}_{{sh}}(z)$ (filled circles).
For $z \geq 1$ the two functions are almost identical. \protect \\
(b) magnification of the appearance-energy curves
for the region $-2 \leq z \leq 4$. Filled circles: shell-corrected
icosahedral values 
[$\Delta E^{{ico}}_{{sh}}(z)
=E^{{ico}}_{{sh}}(z)-E^{{ico}}_{{sh}}(0)$].
Open squares: ETF-DFT values
[$\Delta \widetilde{E}(z)=\widetilde{E}(z)-\widetilde{E}(0)$].
}
\label{fig10}
\end{figure}

\Fref{fig10}(a) shows that the variation of the total ETF-DFT energy 
difference (appearance energies)
$\Delta \widetilde{E}(z)=\widetilde{E}(z)-\widetilde{E}(0)$, as a 
function of excess charge $z$ ($|z|=x$), exhibits a parabolic behavior.
The inset in \fref{fig10}(a) exhibiting the quantity 
\begin{equation}
\widetilde{g}(z) = \frac{\widetilde{E}(z)-\widetilde{E}(0)}{z}
+ \widetilde{A}_1,
\label{gsm}
\end{equation}
plotted versus $z$ (open squares), shows 
a straight line which crosses the zero energy line at $z=1$. As a result
the total ETF-DFT energy has the form,
\begin{equation}
\widetilde{E}(z) = \widetilde{E}(0) + \frac{z(z-1)e^2}{2C}
- \widetilde{A}_1 z.
\label{etfe}
\end{equation}

Equation (\ref{etfe}) indicates that fullerenes behave on the average like a 
capacitor having a capacitance $C$ {\bf (}the second term on the rhs of Eq.\ 
(\ref{etfe}) corresponds to the charging energy of a classical capacitor, corrected 
for the self-interaction of the excess charge \cite{yl1,yl2}{\bf )}. 
We remark that regarding the system as a classical conductor, where the
excess charge accumulates on the outer surface, yields a value 
of $C=8.32$ $a.u.$ (that is the outer radius of the jellium shell).
Naturally, the ETF calculated value for $C$ is somewhat larger because of the
quantal spill-out of the electronic charge density.
Indeed, from the slope of $\widetilde{g}(z)$ we determine \cite{note2}
$C = 8.84$ $a.u$.

A similar plot of the shell-corrected and icosahedrally modified energy
differences $\Delta E^{{ico}}_{{sh}}(z)
=E^{{ico}}_{{sh}}(z)-E^{{ico}}_{{sh}}(0)$
is shown in \fref{fig10}(b) (in the range $-2 \leq z \leq 4$, filled circles).
The function $g^{{ico}}_{{sh}}(z)$,
defined as in Eq.\ (\ref{gsm}) but with the shell-corrected quantities
($\Delta E^{{ico}}_{{sh}}(z)$ and $A^{{ico}}_1$),
is included in the inset to \fref{fig10}(a) (filled circles).
The shift discernible between  $g^{{ico}}_{{sh}}(-1)$
and $g^{{ico}}_{{sh}}(1)$ is approximately 1.7 $eV$, and
originates from the difference of shell effects on the IPs and EAs (see \tref{tbl1}). 
The segments of the curve $g^{{ico}}_{{sh}}(z)$
in the inset of \fref{fig10}(a), corresponding to positively ($z < 0$) and
negatively ($z > 0$) charged states, are again well approximated by straight 
lines, whose slope is close to that found for $\widetilde{g}(z)$.
Consequently, we may approximate the charging energy,
including shell-effects, as follows,
\begin{equation}
E^{{ico}}_{{sh}}(x) = E^{{ico}}_{{sh}}(0)
+ \frac{x(x-1)e^2}{2C} - A^{{ico}}_1 x,
\label{eshneg}
\end{equation}
for {\it negatively\/} charged states, and 
\begin{equation}
E^{{ico}}_{{sh}}(x) = E^{{ico}}_{{sh}}(0)
+ \frac{x(x-1)e^2}{2C} + I^{{ico}}_1 x,
\label{eshpos}
\end{equation}
for {\it positively\/} charged states. Note that without shell-corrections
(i.e., ETF only) $\widetilde{I}_1-\widetilde{A}_1 = e^2/C = 27.2/8.84\; eV 
\approx 3.1 \; eV $, because of the symmetry of Eq.\ (\ref{etfe}) with respect
to $z$, while the shell-corrected quantities are related as 
$I^{{ico}}_1-A^{{ico}}_1 \approx e^2/C + \Delta_{sh}$,
where the shell correction is $\Delta_{sh} \approx 1.55\; eV$
(from \tref{tbl1}, $I^{{ico}}_1-A^{{ico}}_1 \approx 4.65 \; eV$).

Expression (\ref{eshneg}) for the negatively charged states can be rearranged
as follows (energies in units of $eV$),
\begin{equation}
E^{{ico}}_{{sh}}(x)-E^{{ico}}_{{sh}}(0)=
-2.99 + 1.54 (x-1.39)^2,
\label{eshsq}
\end{equation}
in close agreement with the all-electron LDA result of 
Ref.\ \cite{pede}. 

Equations (\ref{eshneg}) and (\ref{eshpos}) can be used to provide simple
analytical approximations for the higher IPs and EAs. Explicitly written,
$A^{{ico}}_x \equiv E^{{ico}}_{{sh}}(x-1) -
E^{{ico}}_{{sh}}(x) = A^{{ico}}_1  - (x-1)e^2/C$ and
$I^{{ico}}_x = I^{{ico}}_1  + (x-1)e^2/C$.
Such expressions have been used previously \cite{wang} with an assumed value 
for $C \approx 6.7\; a.u.$ (i.e., the radius of the C$_{60}$ molecule, as
determined by the distance of carbon nuclei from the center of the 
molecule), which is appreciably smaller than the value obtained by us
($C=8.84 \; a.u.,$ see above) via a microscopic calculation.
Consequently, using the above expression with our calculated value
for $A^{{ico}}_1 = 2.75 \; eV$ (see \tref{tbl1}), we obtain 
an approximate value of $A^{{ico}}_2 = -0.35 \; eV$ (compared to the 
microscopically calculated value of $-0.09\; eV$
given in \tref{tbl1}, and $-0.11\; eV$ obtained by Ref.\ 
\cite{pede}) --- indicating metastability of C$_{60}^{2-}$ ---
while employing an experimental value for 
$A^{{ico}}_1 = 2.74\; eV$, a value of $A^{{ico}}_2 = 0.68\; eV$
was calculated in Ref.\ \cite{wang}. 

Concerning the cations, our expression (\ref{eshpos}) with a calculated
$I^{{ico}}_1 = 7.40\; eV$ (see \tref{tbl1}) and $C=8.84 \; a.u.$ 
yields approximate values 18.5 $eV$ and 31.5 $eV$ for the appearance energies 
of C$_{60}^{2+}$ and C$_{60}^{3+}$ (compared to the microscopic
calculated values of 17.71 $eV$ and 30.99 $eV$, respectively,
extracted from \tref{tbl1}, and 18.6 $eV$ for the former obtained in Ref.\
\cite{rose}). Employing an experimental value for 
$I^{{ico}}_1 = 7.54 \; eV$, corresponding values of 19.20 $eV$ and
34.96 $eV$ were calculated in Ref.\ \cite{wang}. As discussed in Ref.\
\cite{baba}, these last values are rather high, and the origin of the
discrepancy may be traced to the small value of the capacitance which was
used in obtaining these estimates in Ref.\ \cite{wang}.

A negative value of the second affinity indicates that C$_{60}^{2-}$ is
unstable against electron autodetachment. In this context, we note that 
the doubly negatively charged molecule C$_{60}^{2-}$ has been observed in 
the gas phase and is believed to be a long-lived metastable species
\cite {hett,limb}. Indeed, as we discuss
in the next section, the small DFT values of $A^{{ico}}_2$ found by us
and by Ref.\ \cite{pede} yield lifetimes which are much longer than those
estimated by a pseudopotential-like Hartree-Fock model
calculation \cite{hett}, where a value of $\sim$ 1 $\mu s$ was estimated.

\subsubsection{Lifetimes of metastable anions, C$_{60}^{~x-}$}

The second and higher electron affinities of C$_{60}$ were found
to be negative, which implies that the anions C$_{60}^{x-}$ with
$x \geq 2 $ are not stable species, and can lower their
energy by emitting an electron. However, unless the number of excess
electrons is large enough, the emission of an excess electron involves
tunneling  through a barrier. Consequently, the moderately
charged anionic fullerenes can be described as metastable species 
possessing a decay lifetime.

To calculate the lifetime for electron autodetachmant, it is necessary
to determine the proper potential that the emitted electron sees as
it leaves the molecule. The process is analogous to alpha-particle
radioactivity of atomic nuclei. The emitted electron will have a
final kinetic energy equal to the negative of the corresponding higher
EA. We estimate the lifetime of the decay process by using the WKB method,
in the spirit of the theory of alpha-particle radioactivity, which has
established that the main factor in estimating lifetimes is
the relation of the kinetic energy of the emitted particle to the 
Coulombic tail, and not the details of the many-body problem in the 
immediate vicinity of the parent nucleus.

Essential in
this approach is the determination of an appropriate single-particle potential
that describes the transmission barrier. It is well known that the (DFT) LDA
potential posseses the wrong tail, since it allows for the electron to
spuriously interact with itself. A more appropriate potential would be one
produced by the Self-Interaction Correction method of Ref. \cite{perd2}.
This potential has the correct Coulombic tail, but in the case of
the fullerenes presents another drawback, namely Koopman's theorem is not
satisfied to an extent adequate for calculating lifetimes \cite{note3}.
In this context, we note that Koopman's theorem is known to be poorly
satisfied for the case of fullerenes even in Hartree-Fock calculations 
\cite{cios}. Therefore, the HOMO corresponding to the emitted electron,
calculated as described above, cannot be used in the WKB tunneling
calculation.

Since the final energy of the ejected electron equals the negative of
the value of the electron
affinity, we seek a potential that, together with the icosahedral 
perturbation, yields a HOMO level in C$_{60}^{x-}$ with energy 
$-A_x^{{ico}}$.
We construct this potential through a self-interaction correction to the LDA
potential as follows,
\begin{equation}
V_{{WKB}} = V_{{LDA}}[\widetilde{\rho}] - 
V_{H}[\frac{\widetilde{\rho}}{N_e}] -
V_{{xc}}[\xi\frac{\widetilde{ \rho}}{N_e}],
\label{sicxi}
\end{equation}
where the parameter $\xi$ is adjusted so that the HOMO level of C$_{60}^{x-}$
equals $-A_x^{{ico}}$.
In the above expression, the second term on the rhs is an average 
self-interaction Hartree correction which ensures a proper long-range
behavior of the potential (i.e., correct Coulomb tail), and
the third term is a correction to the short-range exchange-correlation.

\begin{figure}[t]
\centering\includegraphics[width=7.cm]{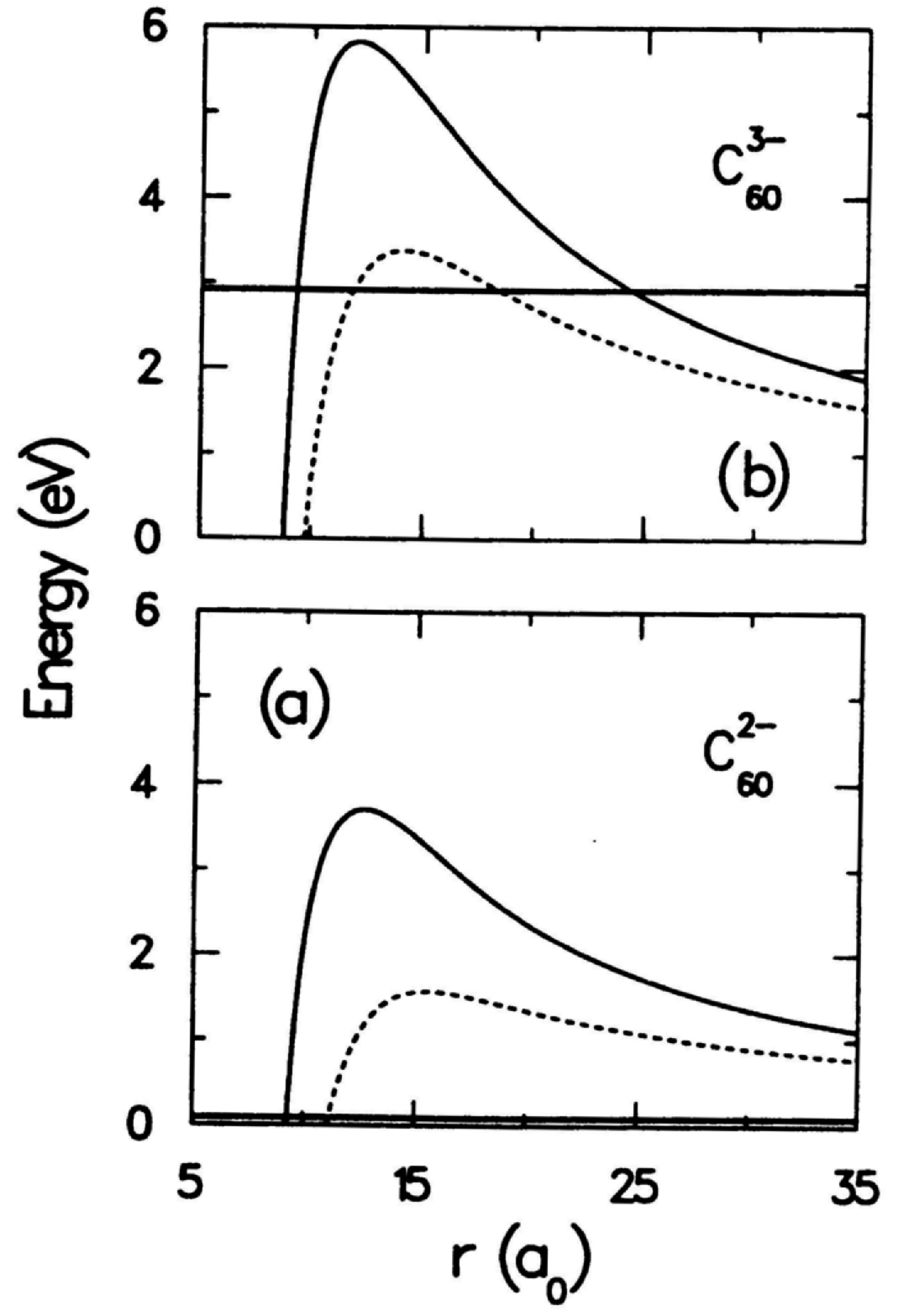}
\caption{
WKB effective barriers used to estimate lifetimes for
C$_{60}^{2-}$ (a) and C$_{60}^{3-}$ (b). Dashed lines correspond to barriers 
due solely to Coulombic repulsion and solid lines to total barriers after adding
the centrifugal components. The thick horizontal solid lines
correspond to the negative of the associated electron
affinities $A^{{ico}}_2$ (a) and  $A^{{ico}}_3$ (b).
In the case of C$_{60}^{2-}$ [panel (a)],
the horizontal solid line at $-A^{{ico}}_2=0.09\; eV$ crosses the total
barrier at an inside point $R_1 = 9.3 \; a.u.$ and again at a distance  
very far from the center of the fullerene molecule, namely at an 
outer point $R_2 = -e^2/A^{{ico}}_2 = 27.2/0.09\; a.u. = 302.2\; a.u.$
This large value of $R_2$, combined with the large centrifugal barrier,
yields a macroscopic lifetime for the metastable C$_{60}^{2-}$
(see text for details).
}
\label{fig11}
\end{figure}

For the cases of C$_{60}^{2-}$ and C$_{60}^{3-}$ such potentials are
plotted in \fref{fig11}. We observe that they have the correct
Coulombic tail, namely a tail corresponding to one electron for
C$_{60}^{2-}$ and to two electrons for C$_{60}^{3-}$.
The actual barrier, however, through which the electron tunnels
is the sum of the Coulombic barrier plus the contribution of
the centrifugal barrier. As seen from \fref{fig11}, 
the latter is significant, since the HOMO in the fullerenes 
possesses a rather high angular momentum ($l=5$), while being
confined in a small volume.

Using the WKB approximation \cite{baz}, we estimate for C$_{60}^{2-}$ a 
macroscopic half-life of $\sim 4 \times 10^{7} $ $years$, while for C$_{60}^{3-}$
we estimate a very short half-life of $2.4 \times 10^{-12}\; s$.
Both these estimates are in correspondence with observations. Indeed,
C$_{60}^{3-}$ has not been observed as a free molecule, while
the free C$_{60}^{2-}$ has been observed to be long lived \cite{hett,limb} 
and was detected even 5 $min$ after its production through
laser vaporization \cite{limb}.

We note that the WKB lifetimes calculated for tunneling 
through Coulombic barriers are very sensitive 
to the final energy of the emitted particle and can vary by many orders of 
magnitude as a result of small changes in this energy, a feature well known
from the alpha radioctivity of nuclei \cite{baz}.

Since the second electron affinity of C$_{60}$ is small, effects due
to geometrical relaxation and spin polarization
can influence its value and, consequently, the estimated lifetime. 
Nevertheless, as shown in Ref.\ \cite{pede}, inclusion of such corrections 
yields again a negative second affinity,
but of somewhat smaller magnitude, resulting in an even longer lifetime
(the sign conventions in Ref.\ \cite{pede} are the opposite of
ours).

Furthermore, as discussed in Ref.\ \cite{coul}, the stabilization effect
of the Jahn-Teller relaxation for the singly-charged ion is only of the
order of 0.03 -- 0.05 $eV$. Since this effect is expected to be largest
for singly-charged species, C$_{60}^{2-}$ is not expected to be 
influenced by it \cite{pede}.

On the other hand, generalized
exchange-correlation functionals with gradient corrections yield slightly 
larger values for the second electron affinity. For example, using
exchange-correlation gradient
corrections, Ref.\ \cite{pede} found $A^{{ico}}_2 = - 0.3 \; eV$,
which is higher (in absolute magnitude) than the value obtained without 
such corrections. This value of $-0.3 \; eV$ leads to a much 
smaller lifetime than the several million of years that correspond to the 
value of $-0.09 \; eV$ calculated by us. Indeed, using the barrier displayed in 
\fref{fig11}(a), we estimate a lifetime for C$_{60}^{2-}$ of approx. 0.37 $s$,
when $A^{{ico}}_2 = - 0.3 \; eV$. 
We stress, however, that even this lower-limit value still corresponds to
macroscopic times and is 5 orders of magnitude larger than the estimate of
Ref.\ \cite{hett}, which found a lifetime of 1 $\mu s$ for 
$A^{{ico}}_2 = - 0.3 \; eV$, since it omitted the large centrifugal 
barrier. Indeed, when we omit the centrifugal barrier, we find a lifetime
estimate of 1.4 $\mu s$, when $A^{{ico}}_2 = - 0.3 \; eV$.
      
\subsection{On mesoscopic forces and quantized conductance in model metallic
nanowires}
\label{wire}

\subsubsection{Background and motivation}

In this section, we show that certain aspects of the mechanical response (i.e.,
elongation force) and electronic transport (e.g., quantized conductance) in
metallic nanowires can be analyzed using the DFT shell 
correction method, developed and applied previously
in studies of metal clusters (see \sref{scm} and \sref{metclu}).
Specifically, we show that in a jellium-modelled,
volume-conserving nanowire, variations of the total energy (particularly 
terms associated with electronic subband corrections) upon elongation of the
wire lead to {\it self-selection\/} of a sequence of stable ``magic''
wire configurations (MWC's, specified in our model by a sequence of the wire's
radii), with the force required to elongate the wire from one configuration to 
the next exhibiting an oscillatory behavior. Moreover, we show that due to the
quantized nature of electronic states in such wires, the electronic conductance
varies in a quantized step-wise manner (in units of the conductance quantum
$g_0=2e^2/h$), correlated with the transitions between MWC's and the 
above-mentioned force oscillations.

Prior to introducing the model, it is appropriate
to briefly review certain previous theoretical and experimental investigations,
which form the background and motivation for this study of nanowires. Atomistic
descriptions, based on realistic interatomic interactions, and/or 
first-principles modelling and simulations played an essential role in
discovering the formation of nanowires, and in predicting and elucidating
the microscopic mechanisms underlying their mechanical, spectral, electronic
and transport properties. 

Formation and mechanical properties of interfacial junctions (in the form of 
crystalline nanowires) have been predicted through early molecular-dynamics 
simulations \cite{land1}, where the materials (gold) were
modelled using semiempirical embedded-atom potentials. In these studies it has
been shown that separation of the contact between materials leads to 
generation of a connective junction which elongates and narrows through
a sequence of structural instabilities; at the early stages, elongation
of the junction involves multiple slip events, while at the later stages,
when the lateral dimension of the wire necks down to a diameter of about 15 
\AA, further elongation involves a succession of stress accumulation and
fast relief stages associated with a sequence of order-disorder structural
transformations localized to the neck region \cite{land1,land2,land3}. These 
structural evolution patterns have been shown through the simulations to be 
portrayed in oscillations of the force required to elongate the wire, with a 
period approximately equal to the interlayer spacing. 
In addition, the ``sawtoothed'' character of the predicted force 
oscillations {\bf (}see Fig.\ 3(b) in Ref.\ \cite{land1} and
Fig.\ 3 in Ref.\ \cite{land2}{\bf )} reflects the stress accumulation and
relief stages of the elongation mechanism. Moreover, the critical resolved yield 
stress of gold nanowires has been predicted \cite{land1,land2} to be $\sim$ 4GPa, 
which is over an order of magnitude larger than that of the bulk, and is comparable 
to the theoretical value for Au (1.5 GPa) in the absence of dislocations.

These predictions, as well as anticipated electronic conductance properties
\cite{land1,boga1}, have been corroborated in a number of experiments using 
scanning tunneling and force microscopy 
\cite{land1,pasc1,oles,pasc2,smith,rubi,stal}, break junctions \cite{krans}, 
and pin-plate techniques \cite{costa,land2} at ambient environments, as well as under 
ultrahigh vacuum and/or cryogenic conditions. Particularly, pertinent
to this section are experimental observations of the oscillatory
behavior of the elongation forces and the correlations between the changes in 
the conductance and the force oscillations; see especially the simultaneous
measurements of force and conductance in gold nanowires in Ref.\ \cite{rubi},
where in addition the predicted ``ideal'' value of the critical yield
stress has also been measured (see also Ref.\ \cite{stal}).

The jellium-based model introduced in this paper,
which by construction is devoid of atomic crystallographic
structure, does not address issues pertaining to nanowire formation methods,
atomistic configurations, and mechnanical response modes {\bf (}e.g., 
plastic deformation mechanisms, interplanar slip, ordering and disordering 
mechanisms (see detailed descriptions in Refs.\ \cite{land1,land2}
and \cite{land3}, and a discussion of conductance dips in Ref.\ \cite{pasc2}),
defects, mechanichal reversibility \cite{rubi,land2}, and roughening of the wires' 
morphology during elongation \cite{land3}{\bf )}, nor does it consider the effects of
the above on the electron spectrum, transport properties, and dynamics \cite{barn5}.
Nevertheless, as shown below, the model offers a useful framework for linking 
investigations of solid-state structures of reduced dimensions (e.g., nanowires) with
methodologies developed in cluster physics, as well as highlighting certain nanowire
phenomena of mesoscopic origins and their analogies to clusters.

\subsubsection{The jellium model for metallic nanowires: Theoretical
method and results}

Consider a cylindrical jellium wire of length $L$, having a positive background 
with a circular cross section of constant radius $R \ll L$ \cite{note45}.
For simplicity, we restrict ourselves here to this symmetry of the wire
cross section. Variations in the shape of the nanowire cross section
serve to affect the degeneracies of the electronic spectrum \cite{sche,boga3} 
without affecting our general conclusions. We also do not include here 
variations of the wire's shape along its axis. Adiabatic variation of
the wire's axial shape introduces a certain amount of smearing of the 
conductance steps through tunnelling, depending on the axial radius of
curvature of the wire \cite{sche,boga2,boga3}. Both the cross-sectional and axial 
shape of the wire can be included in our model in a rather straightforward manner.

As elaborated in \sref{scm}, the principal idea of the SCM is the separation of the
total DFT energy $E_T(R)$ of the nanowire as 
\begin{equation}
E_T(R) = 
\widetilde{E}(R) 
+ \Delta E_{sh} (R),
\label{etscm}
\end{equation}
where $\widetilde{E}(R)$ varies smoothly as a function of the radius $R$ of the wire
(instead of the number of electrons $N$ used in \sref{scm}),
and $\Delta E_{sh} (R)$ is the shell-correction term arising from
the discrete quantized nature of the electronic levels. Again, as elaborated in
\sref{scm}, the smooth contribution in Eq. (\ref{etscm}) is identified with 
$E_{ETF}[\widetilde{\rho}]$. The trial radial lateral density 
$\widetilde{\rho}(r)$ is given by Eq.\ (\ref{rho}), and the constant $\rho_0$ at a 
given radius $R$ is obtained under the normalization condition (charge neutrality)
$2 \pi \int \widetilde{\rho} (r) r dr=
\rho^{(+)}_L(R)$, where $\rho^{(+)}_L (R) = 3R^2/(4r_s^3)$ is the
linear positive background density. Using the optimized 
$\widetilde{\rho}$, one solves for the eigenvalues
$\widetilde{\epsilon}_i$ 
of the Hamiltonian 
$H=-(\hbar^2/2m) \nabla^2 +
V_{ETF}[\widetilde{\rho}]$,
and the shell correction is given by
\begin{eqnarray}
\Delta E_{sh} & \equiv &
E_\mathrm{Harris}[\widetilde{\rho}] -
E_{ETF}[\widetilde{\rho}] \nonumber \\
 & = & \sum_{i=1}^{\mathrm{occ}}
\widetilde{\epsilon}_i -
\int \widetilde{\rho} ({\bf r}) 
V_{ETF}[\widetilde{\rho} ({\bf r})] d{\bf r} -
T_{ETF}[\widetilde{\rho}],
\label{shcor}
\end{eqnarray}
where the summation extends over occupied levels.
Here the dependence of all quantities on the pertinent size variable
(i.e., the radius of the wire $R$) is not shown explicitly. Additionally, the
index $i$ can be both discrete and continuous, and in the latter case the
summation is replaced by an integral (see below).
 
\begin{figure}[t]
\centering\includegraphics[width=7.cm]{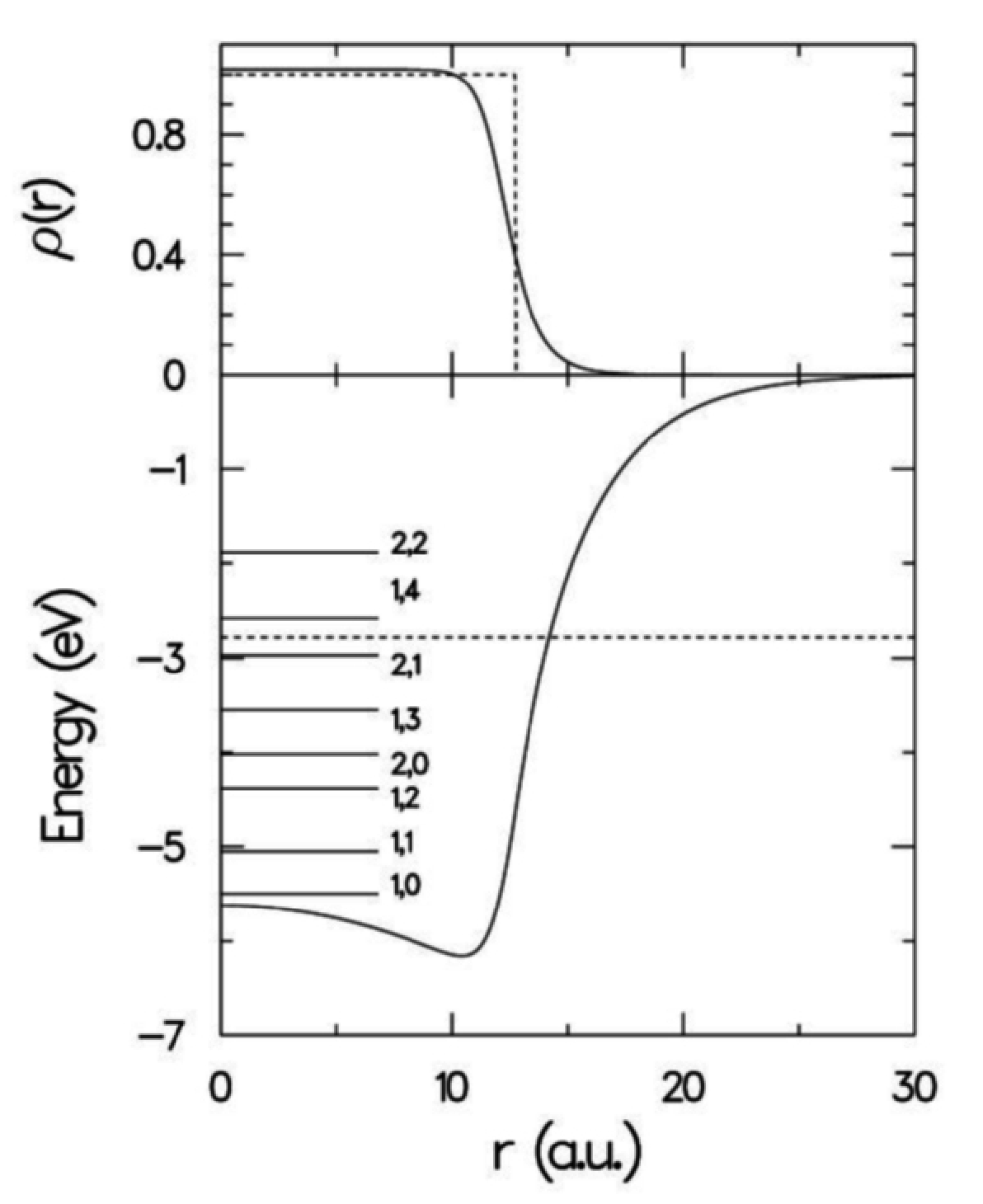}
\caption{Lower panel: The $V_{ETF}(r)$ potential for a sodium wire 
with a uniform jellium background of radius $R=12.7$ a.u., plotted versus
the transverse radial distance from the center of the wire, along with
the locations of the bottoms of the subbands (namely the transverse 
eigenvalues $\widetilde{\epsilon}_{nm}$; $n$ is the number of nodes in the
radial direction plus one, and $m$ is the azimuthal quantum number of the 
angular momentum). The Fermi level is denoted by a dashed line.
Top panel: The jellium background volume density (dashed line) and
the electronic volume density $\widetilde{\rho}(r)$ (solid line, exhibiting
a characteristic spillout) normalized to bulk values are shown.
}
\label{fig12}
\end{figure}

Following the above procedure with a uniform background density of sodium
($r_s=4$ a.u.), a typical potential $V_{ETF}(r)$ for $R=12.7$ a.u., 
where $r$ is the radial coordinate in the transverse plane, is shown in \fref{fig12}, 
along with the transverse eigenvalues $\widetilde{\epsilon}_{nm}$
and the Fermi level; to simplify the calculations of the electronic
spectrum, we have assumed (as noted above) $R \ll L$, which allows us to
express the subband electronic spectrum as 
\begin{equation}
\widetilde{\epsilon}_{nm} (k_z;R) =
\widetilde{\epsilon}_{nm}(R) + \frac{\hbar^2 k_z^2}{2m},
\label{eigval}
\end{equation}
where $k_z$ is the electron wave number along the axis of the wire ($z$).

\begin{figure}[t]
\centering\includegraphics[width=7.cm]{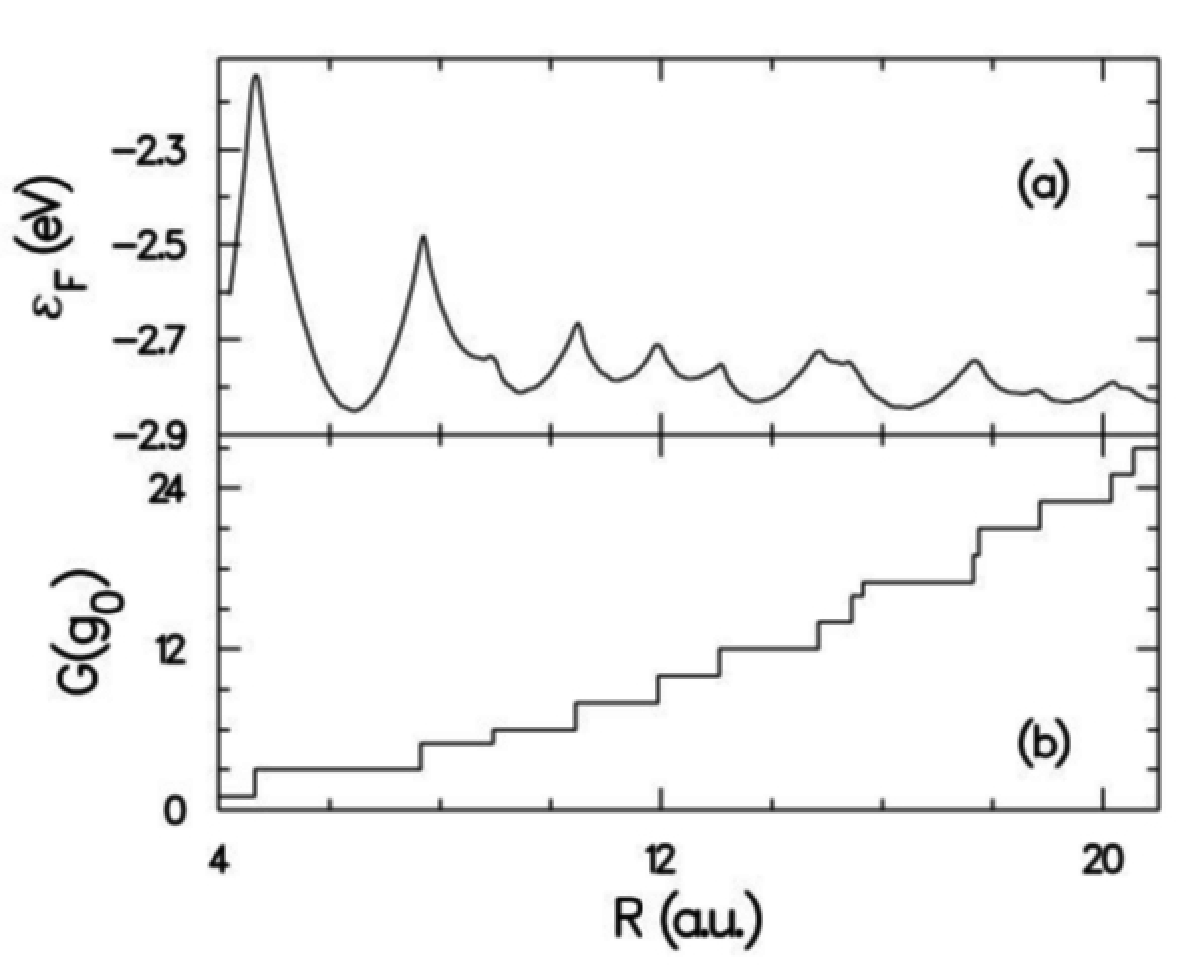}
\caption{
Variation of the Fermi energy $\epsilon_F$ [shown in (a)] and of the
conductance $G$ (shown in (b) 
in units of $g_0=2 e^2/h$), plotted versus the radius R,
for a sodium nanowire. Note the coincidence of the cusps in $\epsilon_F$
with the step-rises of the conductance. The heights of the steps in $G$
reflect the subband degeneracies due to the circular shape of the wire's
cross section.
}
\label{fig13}
\end{figure}

\begin{figure}[t]
\centering\includegraphics[width=7.cm]{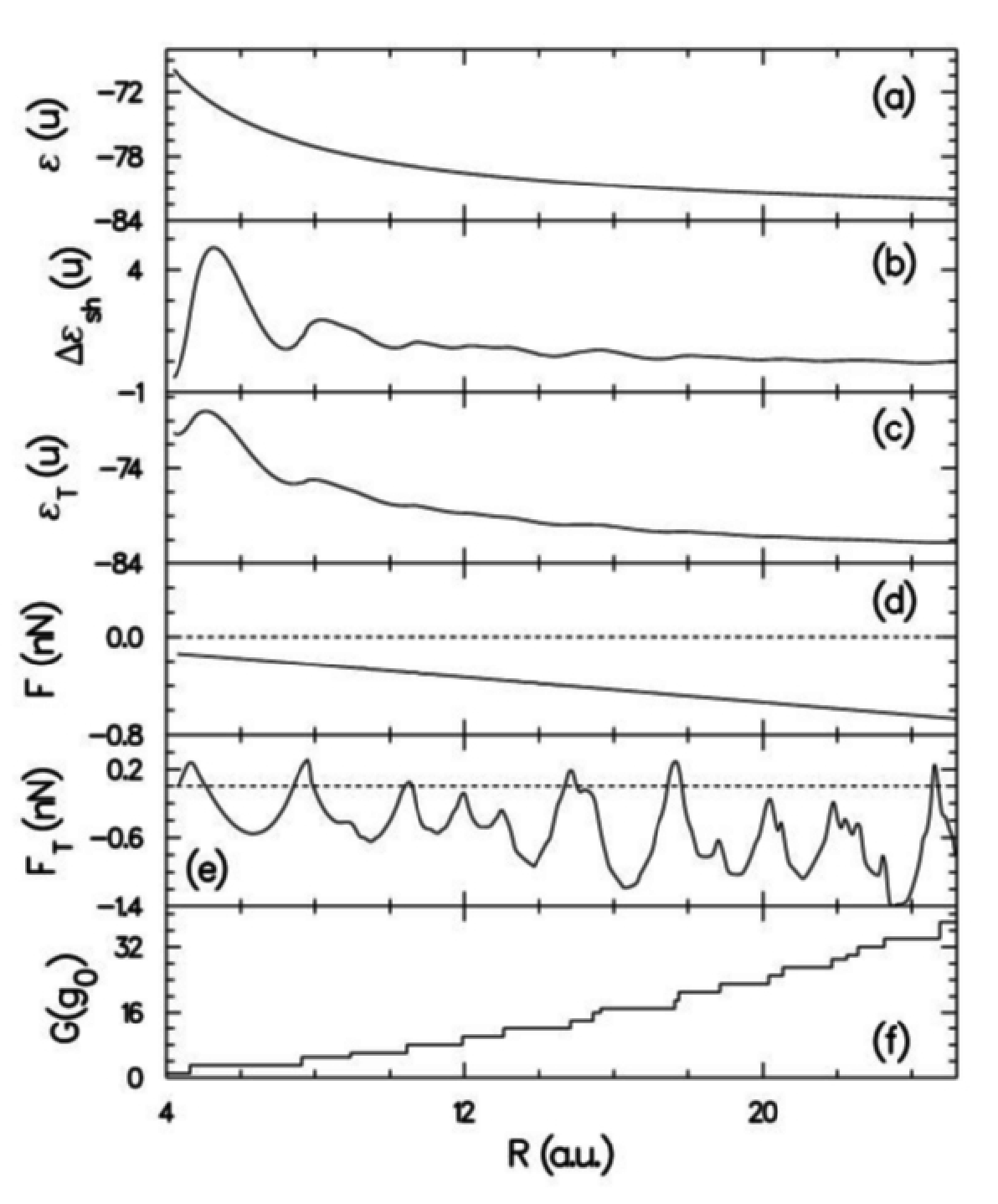}
\caption{
(a-c): The smooth (a) and shell-correction (b) contributions to the total
energy (c) per unit volume of the jellium-modelled sodium nanowire (in units of
$u \equiv 10^{-4}\;$eV/a.u.$^3$), 
plotted versus the radius of the wire (in a.u.). Note the
smaller magnitude of the shell corrections relative to the smooth
contribution. (d-e): The smooth contribution (d) to the total force and
the total force (e), plotted in units of nN versus the wire's radius.
In (e), the zeroes of the force to the left of the force maxima occur at
radii corresponding to the local minima of the energy of the wire (c).
In (f), we reproduce the conductance of the wire (in units of 
$g_0=2e^2/h$), plotted versus R.
Interestingly, calculations of the conductance for the MWC's (i.e.,
the wire radii corresponding to the locations of the step-rises)
through the Sharvin-Weyl formula, \protect\cite{garc,boga3} 
corrected for the finite height of the confining potential \protect\cite{garc} 
(see lower panel of \fref{fig12}), namely
$G=g_0 (\pi S/\lambda_F^2 - \alpha P/\lambda_F$) where $S$ and $P$ are the
area and perimeter of the wire's cross section and $\lambda_F$ is the
Fermi wavelength ($\lambda_F=12.91$ a.u. for Na) with $\alpha=0.1$
(see Ref.\ \protect\cite{garc}), yield results which approximate well the
conductance values (i.e., the values at the bottom of the step-rises)
shown in (f).
}
\label{fig14}
\end{figure}

As indicated earlier,
taking the wire to be charge neutral, the electronic linear density,
$\rho_L^{\mathrm{(--)}}$($R$), must equal the linear positive background density,
$\rho^{(+)}_L (R)$. 
The chemical potential (at $T=0$ the Fermi
energy $\epsilon_F$) for a wire of radius $R$
is determined by setting the expression for the
electronic linear density derived from the subband spectra
equal to $\rho^{(+)}_L (R)$, i.e.,
\begin{equation}
\frac{2}{\pi} 
\sum_{n,m}^{\mathrm{occ}} \sqrt{ \frac{2m}{\hbar^2} [\epsilon_F (R)-
\widetilde{\epsilon}_{nm} (R)] } =
\rho^{(+)}_L (R),
\label{linden}
\end{equation}
where the factor of 2 on the left is due to the spin degeneracy. The summand 
defines the Fermi wave vector for each subband, $k_{F,nm}$. The resulting
variation of $\epsilon_F (R)$ versus $R$ is displayed in \fref{fig13}(a), showing
cusps for values of the radius where a new subband drops below the Fermi
level as $R$ increases (or conversely as a subband moves above the Fermi
level as $R$ decreases upon elongation of the wire).
Using the Landauer expression for the conductance $G$ in the limit of
no mode mixing and assuming unit transmission coefficients, 
$G(R) = g_0 \sum_{n,m} \Theta [\epsilon_F(R)-\widetilde{\epsilon}_{nm}(R)]$, 
where $\Theta$ is the Heaviside step function. The conductance of the nanowire, 
shown in \fref{fig13}(b), exhibits quantized step-wise behavior, with the step-rises 
coinciding with the locations of the cusps in $\epsilon_F (R)$, and the height 
sequence of the steps is 1$g_0$, 2$g_0$, 2$g_0$, 1$g_0$, ..., reflecting the circular
symmetry of the cylindrical wires' cross sections
\cite{boga1}, as observed for sodium nanowires \cite{krans}. Solving for
$\epsilon_F (R)$ [see Eq.\ (\ref{linden})], the expression for the sum on the
right-hand-side of Eq.\ (\ref{shcor}) can be written as
\begin{eqnarray}
&& \sum_i^{\mathrm{occ}} \widetilde{\epsilon}_i  =  
\frac{2}{\pi} \sum_{n,m}^{\mathrm{occ}} 
\int_0^{k_{F,nm}} dk_z \widetilde{\epsilon}_{nm}(k_z;R)  =  
\nonumber \\
&& \frac{2}{3\pi} 
\sum_{n,m}^{\mathrm{occ}}
[\epsilon_F(R) + 2 \widetilde{\epsilon}_{nm}(R)]
\sqrt{ \frac{2m}{\hbar^2} [\epsilon_F(R) - 
\widetilde{\epsilon}_{nm}(R)] }, 
\label{sumi}
\end{eqnarray}
which allows one to evaluate 
$\Delta E_{sh}$ [Eq.\ (\ref{shcor})] for each wire radius $R$. Since the
expression in Eq.\ (\ref{sumi}) gives the energy per unit length, we also 
calculate $E_{ETF}$, $T_{ETF}$, and the volume integral in
the second line of Eq.\ (\ref{shcor}) for cylindrical volumes of unit height.
To convert to energies per unit volume [denoted as $\varepsilon_T (R)$,
$\widetilde{\varepsilon}(R)$, and $\Delta \varepsilon_{sh} (R)$] all
energies are further divided by the wire's cross-sectional area, $\pi R^2$.
The smooth contribution and the shell correction to the wire's energy are
shown respectively in \fref{fig14}(a) and \fref{fig14}(b). The smooth contribution
decreases slowly towards the bulk value ($-$2.25 eV per atom \cite{yl1}). 
On the other hand, the shell corrections are much smaller in magnitude and exhibit an
oscillatory behavior. This oscillatory behavior remains visible in the total energy
[\fref{fig14}(c)] with the local energy minima occurring for values 
$R_{\mathrm{min}}$ corresponding to conductance plateaus. The sequence of 
$R_{\mathrm{min}}$ values defines the MWC's, that is a sequence of wire 
configurations of enhanced stability.

From the expressions for the total energy of the wire [i.e., 
$\Omega \varepsilon_T (R)$, 
where $\Omega = \pi R^2 L$ is the volume of the wire]
and the smooth and shell (subband) contributions to it, we can calculate the
``elongation force'' (EF),
\begin{eqnarray}
F_T (R) & = & -\frac{d[\Omega \varepsilon_T (R)]}{dL} 
= - \Omega \left\{ \frac{d \widetilde{\varepsilon} (R)}{dL} + 
\frac{ d [\Delta \varepsilon_{sh} (R) ]}{dL} 
\right\} \nonumber\\
 & \equiv &   \widetilde{F}(R) + \Delta F_{sh}(R).
\label{force}
\end{eqnarray}
Using the volume conservation, i.e., $d(\pi R^2 L)=0$, these forces can be
written as 
$F_T(R)= (\pi R^3/2) d\varepsilon_T (R)/dR$,
$\widetilde{F}(R)= (\pi R^3/2) d\widetilde{\varepsilon} (R)/dR$, and
$\Delta F_{sh} (R)= 
(\pi R^3/2) d [\Delta \varepsilon_{sh} (R)]/dR$.
$\widetilde{F}(R)$ and $F_T(R)$ are shown in \fref{fig14}(d,e). The 
oscillations in the force resulting from the shell-correction
contributions dominate. In all cases, the radii corresponding
to zeroes of the force situated on the left of the force maxima coincide with
the minima in the potential energy curve of the wire, corresponding to the
MWC's.  Consequently, these forces may be interpreted as guiding the 
self-evolution of the wire toward the MWC's.
Also, all the local maxima in the force occur at the locations of
step-rises in the conductance [reproduced in \fref{fig14}(f)], 
signifying the sequential decrease in the number of
subbands below the Fermi level (conducting channels) as the wire narrows 
(i.e., as it is being elongated). Finally the magnitude of the total forces
is comparable to the measured ones (i.e., in the nN range).

\section{Summary}
\label{summ}

While it was understood rather early that the total energy of nuclei can be 
decomposed into an oscillatory part and one that shows a slow ``smooth'' variation as
a function of size, Strutinsky's seminal contribution \cite{stru} was to calculate
the two parts from different nuclear models: the former from the nuclear shell model
and the latter from the liquid drop model. In particular, the calculation of the 
oscillatory part (shell correction term) was enabled by employing an averaging method
that smeared the single particle spectrum associated with a nuclear model potential.

A semiempirical shell-correction method (referred to as SE-SCM) for metal clusters,
that was developed in close analogy to the original phenomenological Strutinsky 
approach, was presented in the Appendix, along with some applications to triaxial 
deformations and fission barriers of metal clusters.

This chapter reviewed primarily the motivation and theory of a microscopic
shell correction method based on density functional theory (often referred to
as DFT-SCM and originally introduced in Ref.\ \cite{yl1}). In  developing the 
DFT-SCM, we have used for the shell correction term (arising from quantum 
interference effects) a derivation that differs from the Strutinsky methodology 
\cite{stru}. Instead, we have shown \cite{yl1} that the shell correction term can be 
introduced through a kinetic-energy-type density functional [see Eq.\ (\ref{tsh}) 
and Eq.\ (\ref{densh})]. 

The DFT-SCM is computationally advantageous, since
it bypasses the self-consistent iteration cycle of the more familiar KS-DFT.
Indeed, the DFT-SCM energy functional depends only on the single-particle density, 
and thus it belongs to the class of orbital-free DFT methods. Compared to previous
OF-DFT approaches, the DFT-SCM represents an improvement in accuracy. 

Applications of the DFT-SCM to condensed-matter nanostructures, and in
particular metal clusters, fullerenes, and nanowires, were presented in \sref{appl}.

\section*{Acknowledgements}
This research was supported by a grant from the U.S. Department of Energy
(Grant No. FG05-86ER45234).

\begin{appendix}[Semi-empirical shell-correction method (SE-SCM)]
\label{a}

As mentioned above already [see, e.g., \sref{preamble}], rather than proceed 
with the microscopic route, Strutinsky proposed a method for
separation of the total energy into smooth and shell-correction terms 
[see Eq.\ (\ref{enshsm})] based on an averaging procedure.
Accordingly, a smooth part, $\widetilde{E}_{{sp}}$, is extracted out
of the sum of the single-particle energies $\sum_i \widetilde{\varepsilon}_i$
[see Eq.\ (\ref{enstr}), or equivalently Eq.\ (\ref{enhar})
with $\rho^\mathrm{in}$ replaced by $\widetilde{\rho}$ and 
$\varepsilon_i^\mathrm{out}$ by $\widetilde{\varepsilon}_i$]
by averaging them through an appropriate procedure. Usually, but not 
necessarily, one replaces the delta functions in the single-particle
density of states
by gaussians or other appropriate weighting functions. As a result, each
single-particle level is assigned an averaging occupation number
$\widetilde{f}_i$, and the smooth part $\widetilde{E}_{{sp}}$ is
formally written as
\begin{equation}
\widetilde{E}_{{sp}}=
\sum_i \widetilde{\varepsilon}_i \widetilde{f}_i.
\label{espoc}
\end{equation}

Consequently, the Strutinsky shell correction is given by
\begin{equation}
\Delta E_{{sh}}^{{Str}} = \sum_{i=1}^\mathrm{occ} \widetilde{\varepsilon}_i -
\widetilde{E}_{{sp}}.
\label{struav}
\end{equation}

The Strutinsky prescription
(\ref{struav}) has the practical advantage of using only the single-particle
energies $\widetilde{\varepsilon}_i$, and not the smooth density 
$\widetilde{\rho}$. Taking advantage of this, the single-particle energies
can be taken as those of an external potential that empirically approximates
the self-consistent potential of a finite system. In the nuclear case,
an anisotropic three-dimensional harmonic oscillator has been used
successfully to describe the shell-corrections in deformed nuclei.

The single-particle smooth part, $\widetilde{E}_{{sp}}$, however,
is only one component of the total smooth contribution, 
$\widetilde{E}[\widetilde{\rho}]$ ($\widetilde{E}_{{HF}}$ in the Hartree-Fock
energy considered by Strutinsky).
Indeed as can be seen from Eq.\ (\ref{enstr}) 
[or equivalently Eq.\ (\ref{enhar})],
\begin{equation}
E_\mathrm{total} \approx \Delta E_{{sh}}^{{Str}} + 
\widetilde{E}[{\widetilde{\rho}}].
\label{ehfsmsh}
\end{equation}

Strutinsky did not address the question of how to calculate microscopically
the smooth part $\widetilde{E}$ (which necessarily entails
specifying the smooth density ${\widetilde{\rho}}$). Instead he circumvented
this question by substituting for $\widetilde{E}$ the empirical
energies, $E_{{LDM}}$, of the nuclear liquid drop model, namely
he suggested that 
\begin{equation}
E_\mathrm{total} \approx \Delta E_{{sh}}^{{Str}} +
E_{{LDM}}.
\label{enhfld}
\end{equation}

In applications of Eq.\ (\ref{enhfld}),
the single-particle energies involved in the averaging [see Eqs.\ 
(\ref{espoc}) and (\ref{struav})] are commonly obtained as 
solutions of a Schr\"{o}dinger
equation with phenomenological one-body potentials. 
This last approximation has been very successful in describing
fission barriers and properties of strongly deformed nuclei using harmonic
oscillator or Wood-Saxon empirical potentials.

In the following (\sref{a1}), we describe the adaptation of the SE-SCM approach to 
condensed-matter finite systems, and in particular to triaxially deformed metal 
clusters. Moreover (\sref{a2}), we will present several figures illustrating 
applications of the SE-SCM to investigations of the effects of triaxial 
shape-deformations on the properties of metal clusters 
\cite{yl4,yann97,yann00,yann01,yann01.2} 
and to studies of large-scale deformations and 
barriers in fission of charged metal clusters \cite{yl5,yl6,yann02}. We note that 
the SE-SCM has been extended to incorporate electronic entropy effects at finite 
temperatures. This latter extension, referred to as finite-temperature (FT)-SE-SCM 
is not described here, but its theory can be found in Ref.\ \cite{yann97}.  

We mention that, in addition, Strutinsky-type calculations using phenomenological 
potentials have been reported for the case of neutral sodium clusters assuming 
axial symmetry in Refs.\ \cite{bulg,frau,reim,grei08}, and for the case of fission 
in Ref.\ \cite{suga}.

\section{Semiempirical shell-correction method for triaxially deformed 
clusters}
\label{a1}

\subsection{Liquid-drop model for neutral and charged deformed clusters}

For neutral clusters, the liquid-drop model \cite{saun2,suga,brac2} (LDM)
expresses the {\it smooth\/}
part, $\widetilde{E}$, of the total energy as the sum of three contributions,
namely a volume, a surface, and a curvature term, i.e.,
\begin{eqnarray}
\widetilde{E} & = & E_{{vol}}+ E_{{surf}}+E_{{curv}} = 
\nonumber \\
~ & ~ & A_v \int d \tau+ \sigma \int dS + A_c \int dS \kappa,
\label{lqd}
\end{eqnarray}
where $d \tau$ is the volume element and $dS$ is the surface differential
element. The local curvature $\kappa$ is defined by the expression
$\kappa = 0.5 (R^{-1}_{{max}}+R^{-1}_{{min}})$, where
$R_{{max}}$ and $R_{{min}}$ are the two principal radii of
curvature at a local point on the surface of the 
jellium droplet which models the cluster. The corresponding coefficients
can be determined by fitting the extended Thomas-Fermi (ETF)-DFT total energy 
$E_{ETF}[\rho]$ (see \sref{dftscm}) for spherical shapes to
the following parametrized expression as a function of the number, $N$,
of atoms in the cluster \cite{note9},
\begin{equation}
E_{{ETF}}^{{sph}} = \alpha_v N + \alpha_s N^{2/3} + \alpha_c
N^{1/3}.
\label{npar}
\end{equation}

The following expressions relate the coefficients $A_v$, $\sigma$, and
$A_c$ to the corresponding coefficients, ($\alpha$'s), in Eq.\ 
(\ref{npar}),
\begin{equation}
A_v = \frac{3}{4 \pi r_s^3} \alpha_v\; ; \;
\sigma = \frac{1}{4 \pi r_s^2} \alpha_s\; ; \;
A_c = \frac{1}{4 \pi r_s} \alpha_c.
\label{conn}
\end{equation}

In the case of ellipsoidal shapes the areal integral and the integrated
curvature can be expressed in closed analytical form with the help of
the incomplete elliptic integrals ${\cal F} (\psi,k)$ and ${\cal E} (\psi,k)$
of the first and second kind \cite{grad}, respectively. 
Before writing the formulas, we need to introduce some notations. 
Volume conservation must be employed, namely 
\begin{equation}
a^\prime b^\prime c^\prime /R_0^3 = abc =1,
\label{volcons}
\end{equation}
where $R_0$ is the radius of a sphere with the same volume
($R_0=r_sN^{1/3}$ is taken to be the radius of the positive jellium
assuming spherical symmetry), and 
$a=a^\prime/R_0$, etc..., are the dimensionless semi-axes.
The eccentricities are defined through the dimensionless semi-axes as follows
\begin{eqnarray}
e_1^2 & = & 1 - (c/a)^2 \nonumber \\
e_2^2 & = & 1 - (b/a)^2 \nonumber \\
e_3^2 & = & 1 - (c/b)^2. 
\label{ecc}
\end{eqnarray}
The semi-axes are chosen so that
\begin{equation}
a \geq b \geq c.
\label{inqax}
\end{equation}

With the notation $\sin \psi = e_1$, $k_2= e_2/e_1$, and $k_3 = e_3/e_1$,
the relative (with respect to the spherical shape) surface and curvature 
energies are given \cite{hass} by 
\begin{equation}
\frac {E^{{ell}}_{{surf}}}{E^{{sph}}_{{surf}}} =
\frac{ab}{2} \left[ \frac{1-e_1^2}{e_1} {\cal F} (\psi, k_3)
+ e_1 {\cal E}(\psi, k_3) + c^3 \right]
\label{esurf}
\end{equation}
and
\begin{equation}
\frac { E^{{ell}}_{{curv}} } { E^{{sph}}_{{curv}} } =
\frac{bc}{2a} \left[ 1 + \frac{a^3}{e_1} 
\left( ( 1-e_1^2) {\cal F} (\psi, k_2) + e_1^2 {\cal E} (\psi, k_2) \right)
\right].
\label{ecurv}
\end{equation}

The change in the smooth part of the cluster total energy due to the excess
charge $\pm Z$ was already discussed by us for spherical clusters in the
previous section. The result may be summarized as 
\begin{equation}
\Delta \widetilde{E}^{sph} (Z) =
\widetilde{E}^{sph}(Z)-\widetilde{E}^{sph}(0)=  \mp W Z +
\frac{Z(Z \pm 0.25)e^2}{2(R_0+\delta)},
\label{etznp}
\end{equation}
where the upper and lower signs correspond to negatively and positively
charged states, respectively, $W$ is the work function of the metal,
$R_0$ is the radius of the positive jellium assuming spherical symmetry,
and $\delta$ is a spillout-type parameter.

To generalize the above results to an ellipsoidal shape,
$\phi(R_0+\delta)$ $=$ $e^2/(R_0+\delta)$, which is the value of the
potential on the surface of a spherical conductor, needs to be replaced by
the corresponding expression for the potential on the surface of a 
conducting ellipsoid. The final result, normalized to 
the spherical shape, is given by the expression
\begin{equation}
\frac{\Delta \widetilde{E}^{{ell}}  (Z) \pm WZ}
{\Delta \widetilde{E}^{{sph}} (Z) \pm WZ} =
\frac{bc}{e_1} {\cal F}(\psi, k_2),
\label{ecoul}
\end{equation}
where the $\pm$ sign in front of $WZ$ corresponds to negatively and
positively charged clusters, respectively.

\begin{figure}[t]
\centering\includegraphics[width=8.0cm]{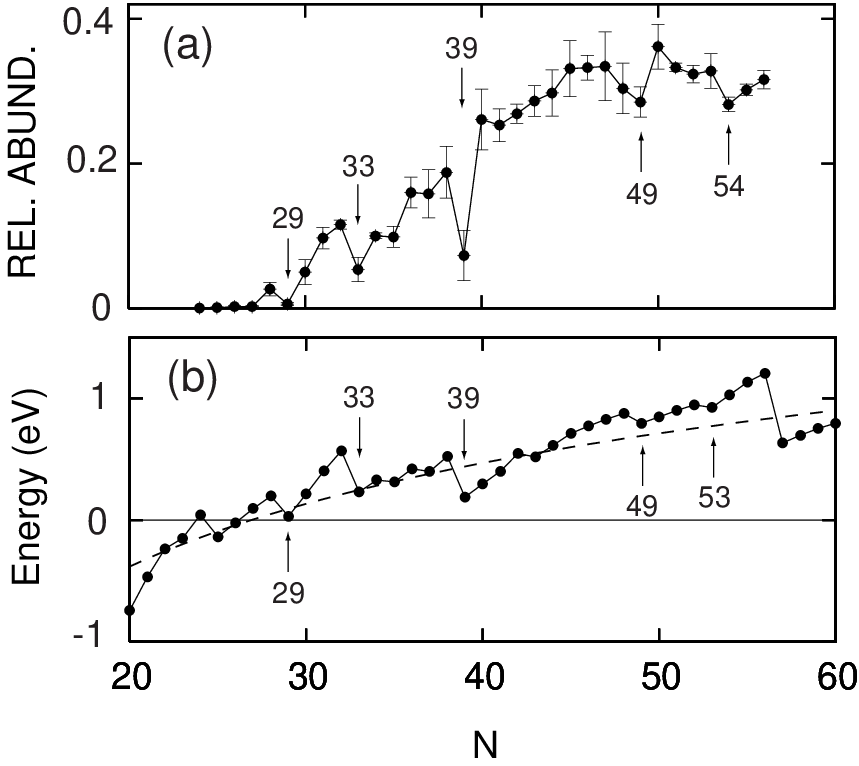}\\
\caption{ 
(a) Experimental yields (denoted as ``REL. ABUND.'') of dianionic silver clusters 
Ag$_N^{2-}$, plotted versus cluster size. The error bars indicate the statistical 
uncertainty. (b) Theoretical FT-SE-SCM \cite{yann97} second electron affinities 
$A_2$ for Ag$_N^{2-}$ clusters at $T=300$ K. LDM results are depicted by the dashed
line. The figure was reproduced from Ref.\ \cite{yann01}. 
}
\label{fig15}
\end{figure}

\subsection{The modified Nilsson potential}

A natural choice for an external potential to be used for calculating
shell corrections with the Strutinsky method is an anisotropic,
three-dimensional oscillator with an ${\bf l}^2$ term for lifting the harmonic
oscillator degeneracies \cite{nils}.
Such an oscillator model for approximating
the total energies of metal clusters, but without separating them into
a smooth and a shell-correction part in the spirit of Strutinsky' s approach,
has been used \cite{dehe2} with some success for calculating relative energy 
surfaces and deformation shapes of metal clusters. 
However, this simple harmonic oscillator model has serious limitations, 
since
i) the total energies are calculated by the expression
$\frac{3}{4} \sum_i
\varepsilon_i$, and thus do not compare with the total energies obtained from
the KS-DFT approach, ii) the model cannot be extended to the case of
charged (cationic or anionic) clusters. Thus absolute ionization potentials,
electron affinities, and fission energetics cannot be calculated in this model.
Alternatively, in our approach, we are making only a limited use of the 
external oscillator potential in calculating a modified Strutinsky shell 
correction. Total energies are evaluated by adding this shell correction to 
the smooth LDM energies. 

In particular, a modified Nilsson Hamiltonian appropriate for metal
clusters \cite{clem,saun} is given by 
\begin{equation}
H_N = H_0 + U_0 \hbar \omega_0 ({\bf l}^2 -< {\bf l}^2>_n),
\label{hn}
\end{equation}
where $H_0$ is the hamiltonian for a three-dimensional anisotropic
oscillator, namely
\begin{eqnarray}
H_0 & = & -\frac{\hbar^2}{2m_e} \bigtriangleup +
       \frac{m_e}{2}(\omega_1^2 x^2 + \omega_2^2 y^2 +\omega_3^2 z^2)
= \nonumber \\
~& ~ & \sum_{k=1}^3 (a_k^\dagger a_k +\frac{1}{2}) \hbar \omega_k.
\label{h0}
\end{eqnarray}

$U_0$ in Eq.\ (\ref{hn}) is a dimensionless parameter, which for occupied
states may depend on the principal quantum number $n=n_1+n_2+n_3$ of the
spherical-oscillator major shell associated with a given level $(n_1,n_2,n_3)$
of the hamiltonian $H_0$ (for clusters comprising up to 100 valence electrons, 
only a weak dependence on $n$ is found, see Table I in Ref.\ \cite{yl4}). 
$U_0$ vanishes for values
of $n$ higher than the corresponding value of the last partially (or fully) 
filled major shell in the spherical limit.

${\bf l}^2=\sum_{k=1}^3 l_k^2$ is a ``stretched'' angular momentum 
which scales to the ellipsoidal shape and is defined as follows,
\begin{equation}
l_3^2 \equiv (q_1p_2 - q_2 p_1)^2,
\label{l3}
\end{equation}
(with similarly obtained expressions for $l_1$ and $l_2$ via a cyclic
permutation of indices) where the stretched position and 
momentum coordinates are defined via the corresponding natural coordinates,
$q^{{nat}}_k$ and $p^{{nat}}_k$, as follows,
\begin{equation}
q_k \equiv q^{{nat}}_k (m_e \omega_k/\hbar)^{1/2} = 
\frac{a_k^\dagger + a_k}{\sqrt{2}}~,~(k=1,2,3),
\label{q}
\end{equation}
\begin{equation}
p_k \equiv p^{{nat}}_k (1 /\hbar m_e \omega_k)^{1/2} = 
i\frac{a_k^\dagger - a_k}{\sqrt{2}}~,~(k=1,2,3).
\label{p}
\end{equation}

\begin{figure}[t]
\centering\includegraphics[width=8.0cm]{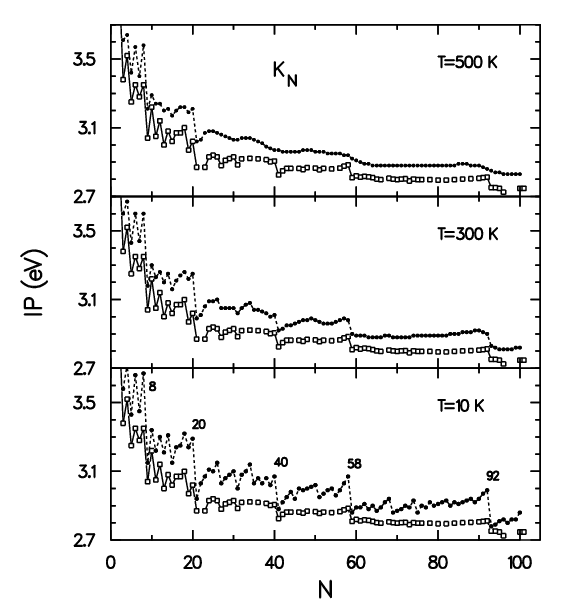}\\
\caption{ 
Ionization potentials of neutral K$_N$ clusters at three temperatures, $T=10$,
300, and 500 K. Solid dots: theoretical FT-SE-SCM \cite{yann97} results. Open
squares: experimental measurements \cite{saun}. The best agreement between
theory and experiment happens for $T=300$ K (room temperature), indicating
the importance of the electronic entropy in quenching the shell effects. 
}
\label{fig16}
\end{figure}

The stretched ${\bf l}^2$ is not a properly defined angular-momentum operator,
but has the advantageous property that it does not mix deformed states which
correspond to sherical major shells with different principal quantum number
$n=n_1+n_2+n_3$ (see, the appendix in Ref.\ \cite{yl4} for the expression of 
the matrix elements of ${\bf l}^2$).

The subtraction of 
the term $<{\bf l}^2>_n =n(n+3)/2$, where $<\; >_n$ denotes the expectation
value taken over the $nth$-major shell in spherical symmetry, 
guaranties that the average separation between major oscillator shells 
is not affected as a result of the lifting of the degeneracy.

The oscillator frequencies can be related to the principal semi-axes $a^\prime$, 
$b^\prime$, and $c^\prime$ [see Eq.\ (\ref{volcons})] via the 
volume-conservation constraint and the requirement that the surface of the
cluster is an equipotential one, namely
\begin{equation}
\omega_1 a^\prime = \omega_2 b^\prime = \omega_3 c^\prime =
\omega_0 R_0,
\label{omeg}
\end{equation}
where the frequency $\omega_0$ for the spherical shape (with radius $R_0$) 
was taken according to Ref.\ \cite{clem2} to be
\begin{equation}
\hbar \omega_0 (N)= 
\frac{49 \; \mbox{eV bohr}^2}{r_s^2 N^{1/3}} 
\left[ 1 + \frac{t}{r_s N^{1/3}} \right]^{-2}.
\label{omeg0}
\end{equation}
Since in this paper we consider solely monovalent elements, $N$ in Eq.\
(\ref{omeg0}) is the number of atoms for the family of clusters
M$_N^{Z \pm}$, $r_s$ is the Wigner-Seitz radius expressed in atomic units, 
and $t$ denotes the electronic spillout for the neutral cluster according
to Ref.\ \cite{clem2}.

\subsection{Averaging of single-particle spectra and semi-empirical
shell correction}

Usually $\widetilde{E}_{{sp}}$ [see Eqs.\ (\ref{espoc}) and (\ref{struav})]
is calculated numerically \cite{nix}. 
However, a variation of the numerical Strutinsky averaging method consists in 
using the semiclassical partition function and in expanding it in powers of
$\hbar^2$. With this method, for the case of an anisotropic, fully triaxial 
oscillator, one finds \cite{bm,bhad} an analytical result, namely
\cite{note33}
\begin{eqnarray}
\widetilde{E}_{{sp}}^{{osc}} & = &
\hbar (\omega_1 \omega_2 \omega_3)^{1/3} \nonumber \\
~ & ~ & \times \left( \frac{1}{4} (3N_e)^{4/3} +
\frac{1}{24} \frac{\omega_1^2 +\omega_2^2 +\omega_3^2}
{(\omega_1 \omega_2 \omega_3)^{2/3}} 
(3N_e)^{2/3} \right),
\label{harmav} 
\end{eqnarray}
where $N_e$ denotes the number of delocalized valence electrons in the cluster.

In the present work, expression (\ref{harmav}) (as modified below) will be 
substituted for the average part
$\widetilde{E}_{{sp}}$ in Eq.\ (\ref{struav}), while the sum
$\sum_i^{\mathrm{occ}} \varepsilon_i$ will be calculated numerically by
specifying the occupied single-particle states of the modified Nilsson
oscillator represented by the hamiltonian (\ref{hn}).

In the case of an isotropic oscillator, not only the smooth contribution, 
$\widetilde{E}_{{sp}}^{{osc}}$, but also the Strutinsky shell 
correction (\ref{struav}) can be specified analytically, \cite{bm}
with the result
\begin{equation}
\Delta E^{{Str}}_{{sh,0}}(x) =
\frac{1}{24} \hbar \omega_0 (3N_e)^{2/3} (-1 + 12 x (1-x)),
\label{harmsh}
\end{equation}
where $x$ is the fractional filling of the highest partially filled 
harmonic oscillator shell. For a filled shell 
($x=0$), $\Delta E^{{Str}}_{{sh,0}}(0) = - \frac{1}{24}
\hbar \omega_0 (3N_e)^{2/3}$, instead of the essentially vanishing value 
as in the case of the ETF-DFT defined shell correction (cf. Fig.\ 1
of Ref.\ \cite{yl4}). To adjust
for this discrepancy, we add $-\Delta E^{{Str}}_{{sh,0}}(0)$ to 
$\Delta E^{{Str}}_{{sh}}$ calculated through Eq.\ (\ref{struav})
for the case of open-shell, as well as closed-shell clusters.

\begin{figure}[t]
\centering\includegraphics[width=11.cm]{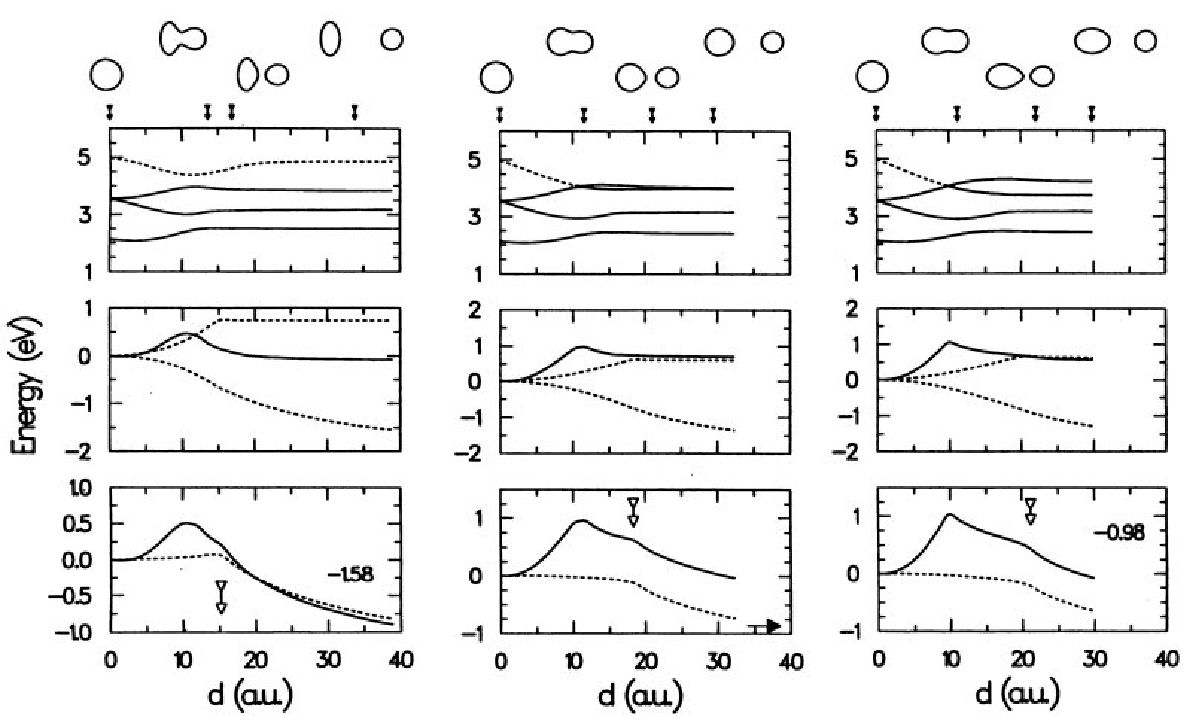}
\caption{
Two-center-oscillator \cite{yl5,yl6} SE-SCM results for the asymmetric channel 
Na$_{10}^{2+}$ $\rightarrow$ Na$_7^+ +$Na$_3^+$. 
The final configuration of Na$_3^+$ is spherical.
For the heavier fragment Na$_7^+$, we present results associated
with three different final shape configurations, namely, oblate 
[(o,s); left], spherical [(s,s); middle], and prolate [(p,s); right]. 
The ratio of shorter over longer
axis is 0.555 for the oblate case and 0.75 for the prolate case. \protect\\
Bottom panel: LDM energy (surface plus Coulomb, dashed curve) 
and total potential energy (LDM plus
shell corrections, solid curve) as a function of fragment separation $d$.
The empty vertical arrow marks the scission point. 
The zero of energy is taken at $d=0$.
A number ($-$1.58 eV or $-$0.98 eV), or a horizontal solid arrow, 
denotes the corresponding dissociation energy.
\protect\\
Middle panel: Shell-correction contribution (solid curve),
surface contribution (upper dashed curve), and Coulomb contribution
(lower dashed curve) to the total energy, as a function of fragment
separation $d$. \protect \\
Top panel: Single-particle spectra as a function of fragment separation
$d$. The occupied (fully or partially) levels are denoted with solid lines. 
The unoccupied levels are denoted with dashed lines. \protect\\
On top of the figure, four snapshots of the evolving cluster shapes are
displayed. The solid vertical arrows mark the corresponding fragment
separations.
Observe that the doorway molecular configurations correspond to the second
snapshot from the left.
Notice the change in energy scale for the middle and bottom panels, as one
passes from (o,s) to (s,s) and (p,s) final configurations.
}
\label{fig17}
\end{figure}

\subsection{Overall procedure}

We are now in a position to summarize the calculational procedure, which 
consists of the following steps:

\begin{enumerate}

\item Parametrize results of ETF-DFT calculations for spherical neutral
      jellia according to Eq.\ (\ref{npar}).

\item Use above parametrization (assuming that parameters per differential 
      element of volume, surface, and integrated curvature are shape 
      independent) in Eq.\ (\ref{lqd}) to calculate the liquid-drop energy 
      associated with neutral clusters, and then add to it the charging
      energy according to Eq.\ (\ref{ecoul}) to determine the total LDM
      energy $\widetilde{E}$.
      
\item Use Equations (\ref{hn}) and (\ref{h0}) for a given deformation
      [i.e., $a^\prime$, $b^\prime$, $c^\prime$, or equivalently
      $\omega_1$, $\omega_2$, $\omega_3$, see Eq.\ (\ref{omeg})] to solve for 
      the single-particle spectrum ($\varepsilon_i$).
      
\item Evaluate the average, 
      $\widetilde{E}_{{sp}}$, of the single-particle spectrum
      according to Eq.\ (\ref{harmav}) and subsequent remarks.

\item Use the results of steps 3 and 4 above to calculate the shell correction
      $\Delta E_{{sh}}^{{Str}}$ according to Eq.\ (\ref{struav}).

\item Finally, calculate the total energy $E_{{sh}}$ as the sum of the
      liquid-drop contribution (step 2) and the shell correction (step 5),
      namely $E_{{sh}}=\widetilde{E}+\Delta E_{{sh}}^{{Str}}$.

\end{enumerate}
      
The optimal ellipsoidal geometries for a given cluster M$^{Z\pm}_N$,
neutral or charged, are determined by systematically varying the distortion 
(namely, the parameters $a$ and $b$) in order to locate the global minimum
of the total energy $E_{{sh}}(N,Z)$.

\section{Applications of SE-SCM to metal clusters}
\label{a2}

As examples of applications of the SE-SCM, we present here three cases. In 
\fref{fig15}, we show experimental electron affinities for doubly 
negatively charged silver clusters \cite{schw01} and compare them 
with theoretical calculations \cite{yann01}. 
In \fref{fig16}, we compare FT-SE-SCM calculations for the IPs of neutral K$_N$ 
clusters with experimental results \cite{yann97}; such comparisons demonstrate the
importance of electronic-entropy effects. Finally, in \fref{fig17}, we display
SE-SCM calculations for the fission barriers associated with the asymmetric channel
Na$_{10}^{2+}$ $\rightarrow$ Na$_7^+ +$Na$_3^+$ \cite{yl5,yl6}; see caption for 
details. The phenomenological binding potential as a function of fission-fragment
separation is described via a two-center-oscillator model \cite{yl5,yl6,grei72}.

A fourth application of the SE-SCM describing the IPs of triaxially deformed cold 
sodium clusters was already used in the introductory \sref{motiv} 
[see \fref{fig1}(c)].

\end{appendix}


\begin{thebibliography}{199}
\bibitem{stru}
V.~M. Strutinsky, 
{\it Nucl. Phys. A\/} {\bf 95\/}, 420 (1967);
{\it Nucl. Phys. A\/} {\bf 122\/}, 1 (1968).
\bibitem{bookring}
P. Ring and P. Schuck,
{\it The Nuclear Many-Body Problem\/}, (Springer, New York, 1980).
\bibitem{yl1}
C. Yannouleas and U. Landman, 
{\it Phys. Rev. B\/} {\bf 48}, 8376 (1993).
\bibitem{tete92}
L.~W. Wang and M.~P. Teter,
{\it Phys. Rev. B\/} {\bf 45}, 13196 (1992).
\bibitem{perr94}
F. Perrot,
{\it J. Phys.: Cond. Matter\/} {\bf 6}, 431 (1994).
\bibitem{madd94}
E. Smargiassi and P.~A. Madden,
{\it Phys. Rev. B\/} {\bf 49}, 5220 (1994).
\bibitem{kaxi05}
T.~J. Frankcombe, G.-J. Kroes, N.~I. Choly, and E. Kaxiras,
{\it J. Phys. Chem. B\/} {\bf 109}, 16554 (2005).
\bibitem{wang00}
Y.~ A. Wang and E.~A. Carter, 
in {\it Theoretical Methods in Condensed Phase Chemistry\/}, S.~D. Schwartz (ed.),
(Kluwer, Dordrecht, 2000), p. 117.
\bibitem{tric09}
S.~B. Trickey, V.~V. Karasiev, and R.~S. Jones,
{\it Int. J. Quantum Chem.\/} {\bf 109}, 2943 (2009).
\bibitem{brack01}
M. Brack,
in {\it Atomic Clusters and Nanoparticles: Les Houches Session LXXIII 
2-28 July 2000\/}, C. Guest, P. Hobza, F. Spiegelman, and F. David (eds.),
(Springer, Berlin, 2001) p. 161.
\bibitem{ullm01}
D. Ullmo, T. Nagano, S. Tomsovic, and H.~U. Baranger, 
{\it Phys. Rev. B\/} {\bf 63}, 125339 (2001). 
\bibitem{delc04}
Ya.~I. Delchev, A.~I. Kuleff, T.~Z. Mineva, F. Zahariev, and J. Maruani,
{\it Int. J. Quantum Chem.\/} {\bf 99}, 265 (2004).
\bibitem{tric04}
W. Zhu, S.~B. Trickey,
{\it Int. J. Quantum Chem.\/} {\bf 100}, 245 (2004). This paper studied 
a perturbative DFT approach in the context of the Harris functional \cite{harr}.
\bibitem{knig84}
W.~D. Knight, K. Clemenger, W.~A. de Heer, W.~A. Saunders, M.~Y. Chou, 
and M.~L. Cohen,
{\it Phys. Rev. Lett.\/} {\bf 52}, 2141 (1984).
\bibitem{deheer93}
W.~A. de Heer, 
{\it Rev. Mod. Phys.\/} {\bf 65}, 611 (1993).
\bibitem{brack93}
M. Brack,
{\it Rev. Mod. Phys.\/} {\bf 65}, 677 (1993).
\bibitem{home}
M.~L. Homer, E.~C. Honea, J.~L. Persson, and R.~L. Whetten, 
(unpublished).
\bibitem{bookekardt}
{\it Metal Clusters\/}, W. Ekardt (ed.), 
(John-Wiley, New York, 1999).
\bibitem{wang07}
B.~J. Zhou and Y.~A. Wang, 
{\it J. Chem. Phys.\/} {\bf 127}, 064101 (2007).
\bibitem{wang08}
B.~J. Zhou and Y.~A. Wang, 
{\it J. Chem. Phys.\/} {\bf 128}, 084101 (2008).
\bibitem{seps} 
U. Landman, R.~N. Barnett, C.~L. Cleveland, and G. Rajagopal,
in {\it Physics and Chemistry of Finite Systems: From Clusters to
Crystals\/}, P. Jena, S.~N. Khanna, and B.~K. Rao (eds.), (Kluwer Academic
Publishers, Dordrecht, 1992), Vol. I, p. 165; J. Jortner, 
{\it Z. Phys. D\/} {\bf 24}, 247 (1992).
\bibitem{pres}
M.~A. Preston and R.~K. Bhaduri, {\it Structure of the Nucleus\/},
(Addison-Wesley, London, 1975).
\bibitem{yl2}
C. Yannouleas and U. Landman, 
{\it Chem. Phys. Lett.\/} {\bf 210}, 437 (1993).
\bibitem{yl3}
R.~N. Barnett, C. Yannouleas, and U. Landman, 
{\it Z. Phys. D\/} {\bf 26}, 119 (1993).
\bibitem{yl4}
C. Yannouleas and U. Landman,  
{\it Phys. Rev. B\/} {\bf 51}, 1902 (1995).
\bibitem{yl5}
C. Yannouleas, R.~N. Barnett, and U. Landman,  
{\it Comments At. Mol. Phys.\/} {\bf 31}, 445 (1995).
\bibitem{yl6}
C. Yannouleas and U. Landman, 
{\it J. Phys. Chem.\/} {\bf 99}, 14577 (1995). 
\bibitem{suga}
H. Koizumi, S. Sugano, and Y. Ishii, 
{\it Z. Phys. D\/} {\bf 28}, 223 (1993);
M. Nakamura, Y. Ishii, A. Tamura, and S. Sugano, 
{\it Phys. Rev. A\/} {\bf 42}, 2267 (1990).
\bibitem{bulg}
A. Bulgac and C. Lewenkopf, 
{\it Phys. Rev. Lett.\/} {\bf 71}, 4130 (1993).
\bibitem{frau}
S. Frauendorf and V.~V. Pashkevich, 
{\it Z. Phys. D\/} {\bf 26}, S 98 (1993).
\bibitem{reim}
S.~M. Reimann, M. Brack, and K. Hansen,
{\it Z. Phys. D\/} {\bf 28}, 235 (1993).
\bibitem{yann97.3}
C. Yannouleas and U. Landman, 
{\it J. Chem. Phys.\/}  {\bf 107}, 1032 (1997). 
\bibitem{yann97} 
C. Yannouleas and U. Landman, 
{\it Phys. Rev. Lett.\/} {\bf 78}, 1424 (1997).
\bibitem{yann99}
C. Yannouleas, U. Landman, and R.~N. Barnett,
{\it Dissociation, Fragmentation and Fission of Simple Metal Clusters\/},
p. 145 in Ref.\ \cite{bookekardt}.
\bibitem{yann00}
C. Yannouleas and U. Landman, 
{\it Rhys. Rev. B\/} {\bf 61}, R10587 (2000).
\bibitem{yann01}
C. Yannouleas, U. Landman, A. Herlert, and L. Schweikhard,
{\it Phys. Rev. Lett.\/}  {\bf 86}, 2996 (2001).
\bibitem{yann01.2}
C. Yannouleas, U. Landman, A. Herlert, and L. Schweikhard,
{\it Eur. Phys. J. D\/}  {\bf 16}, 81 (2001).
\bibitem{yann02}
C. Yannouleas, U. Landman, C. Brechignac, Ph. Cahuzac, B. Concina, 
and J. Leygnier,
{\it Phys. Rev. Lett.\/}  {\bf 89}, 173403 (2002).
\bibitem{yl7}
C. Yannouleas and U. Landman, 
{\it Chem. Phys. Lett.\/} {\bf 217}, 175 (1994).
\bibitem{yann97.2}
C. Yannouleas and U. Landman, 
{\it J. Phys. Chem. B\/} {\bf 101}, 5780 (1997).
\bibitem{yann98}
C. Yannouleas, E.~N. Bogachek, and U. Landman,
{\it Phys. Rev. B\/}  {\bf 57}, 4872 (1998).
\bibitem{staff97}
A. Stafford, D. Baeriswyl, and J. B\"{u}rki, 
{\it Phys. Rev. Lett.\/} {\bf 79}, 2863 (1997). 
\bibitem{bm}
\AA. Bohr and B.~R. Mottelson,  
{\it Nuclear Structure\/} (Benjamin, Reading, Massachusetts, 1975),
Vol. II.
\bibitem{weiz}
C.~F. Von Weizs\"{a}cker, 
{\it Z. Phys.\/} {\bf 96}, 431 (1935).
\bibitem{beth}
H.~A. Bethe and R.~F. Bacher, 
{\it Rev. Mod. Phys.\/} {\bf 8}, 82 (1936).
\bibitem{mysw}
W.~D. Myers and W.~J. Swiatecki,
{\it Nucl. Phys.\/} {\bf 81}, 1 (1966).
\bibitem{snso}
D.~R. Snider and R.~S. Sorbello,
{\it Solid State Commun.\/} {\bf 47}, 845 (1983).
\bibitem{brac2}
M. Brack, 
{\it Phys. Rev. B\/} {\bf 39}, 3533 (1989).
\bibitem{serr}
Ll. Serra, F. Garc\'{i}as, M. Barranco, J. Navarro, L.~C. Balb\'{a}s, 
and A. Ma\~{n}anes, {\it Phys. Rev. B\/} {\bf 39}, 8247 (1989).
\bibitem{memb}
M. Membrado, A.~F. Pacheco, and J. San\~{u}do, 
{\it Phys. Rev. B\/} {\bf 41}, 5643 (1990).
\bibitem{enge}
E. Engel and J.~P. Perdew,
{\it Phys. Rev. B\/} {\bf 43}, 1331 (1991).
\bibitem{seid}
M. Seidl, K.-H. Meiwes-Broer, and M. Brack, 
{\it J. Chem. Phys.\/} {\bf 95}, 1295 (1991).
\bibitem{yann}
C. Yannouleas, R.~A. Broglia, M. Brack, and P.~F. Bortignon, 
{\it Phys. Rev. Lett.\/} {\bf 63}, 255 (1989).
\bibitem{yann1}
C. Yannouleas and R.~A. Broglia,  
{\it Phys. Rev. A\/} {\bf 44}, 5793 (1991);
{\it Europhys. Lett.\/} {\bf 15}, 843 (1991);
C. Yannouleas, P. Jena, and S.~N. Khanna,  
{\it Phys. Rev. B\/} {\bf 46}, 9751 (1992).
\bibitem{yann2}
C. Yannouleas,  
{\it Chem. Phys. Lett.\/} {\bf 193}, 587 (1992).
\bibitem{yann3}
C. Yannouleas and R.~A. Broglia,  
{\it Ann. Phys.\/} (N.Y.) {\bf 217}, 105 (1992);
C. Yannouleas, E. Vigezzi, and R.~A. Broglia,
{\it Phys. Rev. B\/} {\bf 47}, 9849 (1993);
C. Yannouleas, F. Catara, and N. Van Giai,
{\it Phys. Rev. B\/} {\bf 51}, 4569 (1995).
\bibitem{ks}
W. Kohn and L.~J. Sham,  
{\it Phys. Rev.\/} {\bf 140}, A1133 (1965).
\bibitem{dehe2}
W.~A. De Heer,
{\it Rev. Mod. Phys.\/} {\bf 65}, 611 (1993).
\bibitem{kres}
V.~V. Kresin, 
{\it Phys. Rep.\/} {\bf 220}, 1 (1992).
\bibitem{ekar1}
W. Ekardt, W. 
{\it Phys. Rev. B\/} {\bf 31}, 6360 (1985).
\bibitem{beck2}
D.~E. Beck,
(1991) {\it Phys. Rev. B\/} {\bf 43}, 7301 (1991).
\bibitem{harr}
J. Harris, 
{\it Phys. Rev. B\/} {\bf 31}, 1770 (1985).
\bibitem{finn}
M.~W. Finnis,
{\it J. Phys.: Condens. Matter\/} {\bf 2}, 331 (1990).
\bibitem{pome}
H.~M. Polatoglou and M. Methfessel, 
{\it Phys. Rev. B\/} {\bf 37}, 10403 (1988).
\bibitem{foul}
W.~M.~C. Foulkes and R. Haydock, 
{\it Phys. Rev. B\/} {\bf 39}, 12520 (1989).
\bibitem{zare}
E. Zaremba, 
{\it J. Phys.: Condens. Matter\/} {\bf 2}, 2479 (1990).
\bibitem{tf}
L.~H. Thomas,
{\it Proc. Cambridge Philos. Soc.\/} {\bf 23}, 542 (1926);
E. Fermi,   
{\it Z. Phys.\/} {\bf 48}, 73 (1928).
\bibitem{hodg}
C.~H. Hodges,  
{\it Can. J. Phys.\/} {\bf 51}, 1428 (1973).
\bibitem{ekar2}
W. Ekardt,  
{\it Phys. Rev. B\/} {\bf 29}, 1558 (1894).
\bibitem{dehe}
W.~A. De Heer, W.~D. Knight, M.~Y. Chou, and M.~L. Cohen,  
{\it Solid State Phys.\/} {\bf 40}, 93 (1987).
\bibitem{barn}
R.~N. Barnett, U. Landman, and G. Rajagopal,  
{\it Phys. Rev. Lett.\/} {\bf 67}, 3058 (1991);
see also R.~N. Barnett and U. Landman,  
{\it ibid.\/} {\bf 69}, 1472 (1992);
R.~N. Barnett, U. Landman, A. Nitzan, and G. Rajagopal, 
{\it J. Chem. Phys.\/} {\bf 94}, 608 (1991);
H.-P. Cheng, R.~N. Barnett, and U. Landman, 
{\it Phys. Rev. B\/} {\bf 48}, 1820 (1993).
\bibitem{andr}
U. R\"{o}thlisberger and W. Andreoni,  
{\it J. Chem. Phys.\/} {\bf 94}, 8129 (1991).
\bibitem{perd}
J.~P. Perdew and A. Zunger, 
{\it Phys. Rev. B\/} {\bf 23}, 5048 (1981).
\bibitem{alon}
L.~C. Balb\'{a}s, A. Rubio, and J.~A. Alonso,  
{\it Chemical Phys.\/} {\bf 120}, 239 (1988).
\bibitem{ekar8}
Z. Penzar and W. Ekardt, 
{\it Z. Phys. D\/} {\bf 17}, 69 (1990).
\bibitem{wood}
D.~M. Wood, 
{\it Phys. Rev. Lett.\/} {\bf 46}, 749 (1981).
\bibitem{stav}
M.~P.~J. Van Staveren, H.~B. Brom, L.~J. de Jongh, and Y. Ishii, 
{\it Phys. Rev. B\/} {\bf 35}, 7749 (1987).
\bibitem{perd2}
J.~P. Perdew and Y. Wang, 
{\it Phys. Rev. B\/} {\bf 38}, 12228 (1988).
\bibitem{siem}
Ph.~J. Siemens and A.~S. Jensen, 
{\it Elements of nuclei\/} (Addison-Wesley, New York, 1987).
\bibitem{note}
We emphasize that while the effective potentials are significantly different 
when SIC is used, other quantities, such as the total energy,
IPs, and EAs are only slightly altered by SIC as shown in
Ref.\ \cite{perd}, and by our own calculations.
\bibitem{hofm}
S. Hofmann,  
{\it Proton radioactivity\/} in {\it Particle Emission from Nuclei\/},
D.~N. Poenaru and M.~S. Ivascu (eds.), (CRC Press, Boca Raton, Florida, 1989)
Vol. II, p. 25.
\bibitem{brec}
C. Br\'{e}chignac, Ph. Cahuzac, F. Carlier, and J. Leygnier,  
{\it Phys. Rev. Lett.\/} {\bf 63}, 1368 (1989).
\bibitem{trou}
N. Troullier and J.~L. Martins, 
{\it Phys. Rev. B\/} {\bf 46}, 1754 (1992);
J.~L. Martins, N. Troullier, and J.~H. Weaver,  
{\it Chem. Phys. Lett.\/} {\bf 180}, 457 (1991).
\bibitem{koha}
J. Kohanoff, W. Andreoni, and M. Parrinello, 
{\it Chem. Phys. Lett.\/} {\bf 198}, 472 (1992).
\bibitem{yaba}
K. Yabana and G.~F. Bertsch, 
{\it Physica Scripta\/} {\bf 48}, 633 (1993).
\bibitem{lipp}
N. Van Giai and E. Lipparini, 
{\it Z. Phys. D\/} {\bf 27}, 193 (1993).
\bibitem{pusk}
M.~J. Puska and R.~M. Nieminen, 
{\it Phys. Rev. A\/} {\bf 47}, 1181 (1993).
\bibitem{gram}
B. Grammaticos,  
{\it Z. Phys. A\/} {\bf 305}, 257 (1982).
\bibitem{gall}
G.~A. Gallup,  
{\it Chem. Phys. Lett.\/} {\bf 187}, 187 (1991).
\bibitem{hadd}
R.~C. Haddon, L.~E. Brus, and K. Raghavachari, 
{\it Chem. Phys. Lett.\/} {\bf 125}, 459 (1986).
\bibitem{gerl}
M. Gerloch and R.~C. Slade,
{\it Ligand field parameters\/}, (Cambridge Univ. Press, London, 1973).
\bibitem{rose}
A. Ros\'{e}n and B. W\"{a}stberg,
{\it J. Chem. Phys.\/} {\bf 90}, 2525 (1989);
B. W\"{a}stberg and A. Ros\'{e}n,
{\it Physica Scripta\/} {\bf 44}, 276 (1991).
\bibitem{ye}
L. Ye and A.~J. Freeman, 
{\it Chem. Phys.\/} {\bf 160}, 415 (1992).
\bibitem{note2}
Due to the changing spill-out with excess charge $z$,
the capacitance should be written as $C+\delta(z)$. For our
purposes here the small correction $\delta(z)$ can be neglected.
\bibitem{pede}
M.~R. Pederson and A.~A. Quong, 
{\it Phys. Rev. B\/} {\bf 46}, 13584 (1992).
\bibitem{wang}
Y. Wang, D. Tom\'{a}nek, G.~F. Bertsch, and R.~S. Ruoff,  
{\it Phys. Rev. B\/} {\bf 47}, 6711 (1993).
\bibitem{baba}
M. Sai Baba, T.~S. Lakshmi Narasimhan, R. Balasubramanian, and
C.~K. Mathews,  
{\it Int. J. Mass Spectrom. Ion Processes\/} {\bf 125}, R1 (1993).
\bibitem{hett}
R.~L. Hettich, R.~N. Compton, and R.~H. Ritchie, 
{\it Phys. Rev. Lett.\/} {\bf 67}, 1242 (1991).
\bibitem{limb}
P.~A. Limbach, L. Schweikhard, K.~A. Cowen, M.~T. McDermott, 
A.~G. Marshall, and J.~V. Coe,  
{\it J. Am. Chem. Soc.\/} {\bf 113}, 6795 (1991).
\bibitem{note3}
For certain systems, such as for example sodium clusters, an 
orbitally-averaged-like SIC treatment yielded 
highest-occupied-molecular-orbital (HOMO) energies for anions in adequate 
agreement with the calculated electron affinities (see Refs.\ \cite{yl1,yl2}).
\bibitem{cios}
J. Cioslowski and K. Raghavachari,  
{\it J. Chem. Phys.\/} {\bf 98}, 8734 (1993).
\bibitem{baz}
A.~I. Baz', Y.~B. Zel'dovich, and A.~M. Perelomov,
{\it Scattering, reactions, and decay in nonrelativistic quantum mechanics\/},
(Israel Program for Scientific Translations Ltd., Jerusalem, 1969).
\bibitem{coul}
V. De Coulon, J.~L. Martins, and F. Reuse,  
{\it Phys. Rev. B\/} {\bf 45}, 13 671 (1992).
\bibitem{land1}
U. Landman,  
W.~D. Luedtke,
N. Burnham,
and R.~J. Colton,
{\it Science\/} {\bf 248}, 454 (1990).
\bibitem{land2}
U. Landman, 
W.~D. Luedtke, 
B.~E. Salisbury,
and R.~.L. Whetten,
{\it Phys. Rev. Lett.\/} {\bf 77}, 1362 (1996).
\bibitem{land3}
U. Landman, 
W.~D. Luedtke,
and J. Gao,
{\it Langmuir\/} {\bf 12}, 4514 (1996).
\bibitem{boga1}
E.~N. Bogachek,
A.~M. Zagoskin,
and I.~O. Kulik,
{\it Fiz. Nizk. Temp.\/} {\bf 16}, 1404 (1990) 
[{\it Sov. J. Low Temp. Phys.} {\bf 16}, 796 (1990)].
\bibitem{pasc1}
J.~I. Pascual,
J. Mendez,
J. Gomez-Herrero,
J.~M. Baro,
N. Garcia,
and V.~T. Binh,
{\it Phys. Rev. Lett.} {\bf 71}, 1852 (1993).
\bibitem{oles}
L. Olesen,
E. Laegsgaard,
I. Stensgaard, 
F. Besenbacher, 
J. Schiotz, 
P. Stoltze,
K.~W. Jacobsen,
and J.~N. Norskov,
{\it Phys. Rev. Lett.\/} {\bf 72}, 2251 (1994).
\bibitem{pasc2}
J.~I. Pascual,
J. Mendez,
J. Gomez-Herrero,
J.~M. Baro,
N. Garcia,
U. Landman, 
W.~D. Luedtke,
E.~N. Bogachek,
and H.-P. Cheng,
{\it Science\/} {\bf 267}, 1793 (1995).
\bibitem{smith}
D.~P.~E. Smith, 
{\it Science\/} {\bf 269}, 371 (1995).
\bibitem{rubi}
G. Rubio, 
N. Agrait,
and S. Vieira,
{\it Phys. Rev. Lett.\/} {\bf 76}, 2302 (1996).
\bibitem{stal}
A. Stalder and U. Durig,
{\it Appl. Phys. Lett.\/} {\bf 68}, 637 (1996).
\bibitem{krans}
J.~M. Krans,
J.~M. van Ruitenbeek,
V.~V. Fisun,
I.~K. Yanson,
and L.~J. de Jongh,
{\it Nature\/} {\bf 375}, 767 (1995).
\bibitem{costa}
J.~L. Costa-Kramer, 
N. Garcia,
P. Garcia-Mochales, 
and P.~A. Serena, 
{\it Surface Science\/} {\bf342}, 11144 (1995).
\bibitem{barn5}
R.~N. Barnett and U. Landman,
{\it Nature\/} {\bf 387}, 788 (1997).
\bibitem{note45}
For an axially symmetric nanowire with variable radius, see Ref.\ \cite{yann98}.
\bibitem{sche}
A.~G. Scherbakov,
E.~N. Bogachek,
and U. Landman,
{\it Phys. Rev. B\/} {\bf 53}, 4054 (1996).
\bibitem{boga3}
E.~N. Bogachek,
A.~G. Scherbakov,
and U. Landman, 
{\it Phys. Rev. B\/} {\bf 56}, 1065 (1997). 
\bibitem{boga2}
E.~N. Bogachek,
A.~G. Scherbakov,
and U. Landman,
{\it Phys. Rev. B\/} {\bf 53}, R13246 (1996).
\bibitem{garc}
A. Garcia-Martin, 
J.~A. Torres,
and J.~J. Saenz,
{\it Phys. Rev. B\/} {\bf 54}, 13448 (1996).
\bibitem{grei08}
R.~A. Gherghescu, D.~N. Poenaru, A. Solov'yov, and W. Greiner,
{\it Int. J. Mod. Phys. B\/} {\bf 22}, 4917 (2008). 
\bibitem{saun2}
W.~A. Saunders,
{\it Phys. Rev. A\/} {\bf 46}, 7028 (1992).
\bibitem{note9}
Here, we consider clusters of monovalent elements (Na, K, and Cu).
For polyvalent elements, $N$ in Eq.\ (\ref{npar}) must be replaced by $Nv$,
where $v$ is the valency.
\bibitem{grad}
I.~S. Gradshteyn and I.~M. Ryzhik,
{\it Table of integrals, series, and products\/},
(Academic, New York, 1980) Ch. 8.11.
\bibitem{hass}
R.~W. Hasse and W.~D. Myers, 
{\it Geometrical relationships of macroscopic nuclear physics\/},
(Springer-Verlag, Berlin, 1988) Ch. 6.5.
\bibitem{nils}
S.~G. Nilsson,
{\it K. Danske Vidensk. Selsk. Mat.-Fys. Medd.\/} {\bf 29}, No. 16 (1955).
\bibitem{clem}
K.~L. Clemenger,
{\it Phys. Rev. B\/} {\bf 32}, 1359 (1985).
\bibitem{saun}
W.~A. Saunders, 
{\it Ph.D. dissertation\/}, University of California, Berkeley (1986);
W.~A. Saunders, K. Clemenger, W.~A. de Heer, and W.~D. Knight,
{\it Phys. Rev. B\/} {\bf 32}, 1366 (1985).
\bibitem{clem2}
K.~L. Clemenger, 
{\it Ph.D. dissertation\/}, University of California, Berkeley (1985).
\bibitem{nix}
J.~R. Nix, 
{\it Annu. Rev. Nucl. Part. Sci.\/} {\bf 22}, 65 (1972).
\bibitem{bhad}
R.~K. Bhaduri and C.~K. Ross, 
{\it Phys. Rev. Lett.\/} {\bf 27}, 606 (1971).
\bibitem{note33}
The perturbation ${\bf l}^2-<{\bf l}^2>_n$ in the hamiltonian (\ref{hn})
influences the shell correction $\Delta E_{{sh}}^{{Str}}$,
but not the average, $\widetilde{E}_{{sp}}$, of the
single-particle spectrum, since $U_0=0$ for all shells with principal quantum
number $n$ higher than the minimum number required for accomodating $N_e$
electrons (see, Ref.\ \cite{bm}, p. 598 ff.).
\bibitem{schw01}
A. Herlert, L. Schweikhard, and M. Vogel,
{\it Eur. Phys. J. D\/} {\bf 16}, 65 (2001). 
\bibitem{grei72}
J. Maruhn and W. Greiner, 
{\it Z. Phys.\/} {\bf 251}, 431 (1972).
\end{thebibliography}
\end{document}